\documentclass[aps,nofootinbib,superscriptaddress, showpacs,preprintnumbers,  nofootinbibt,twocolumn]{revtex4-2}

\usepackage{epsfig}
\usepackage{multirow}
\usepackage{eurosym}
\usepackage{dcolumn}
\usepackage{bm}
\usepackage{enumerate}
\usepackage{float}
\usepackage{epstopdf}
\usepackage{amsmath}
\usepackage{bm}
\usepackage{amsfonts}
\usepackage{amssymb}
\usepackage{graphicx}
\usepackage{alphalph,mathtools}
\usepackage{etoolbox}
\usepackage{color}
\usepackage{booktabs}
\usepackage{footnote}
\usepackage{makecell,tabularx}
\usepackage{subfigure, rotating, bm, array}
\usepackage[export]{adjustbox}
\usepackage{mathtools}
\usepackage{comment}
\usepackage{hyperref}
\usepackage{enumerate}
\usepackage{orcidlink}

\hypersetup{colorlinks,citecolor=blue}
\hypersetup{colorlinks=true,linkcolor=magenta,filecolor=magenta,    urlcolor=blue}

\def\be{\begin{equation}}
\def\ee{\end{equation}}
\def\bea{\begin{eqnarray}}
\def\eea{\end{eqnarray}}

\begin{document}

\title{Shadow and strong gravitational lensing of new wormhole solutions supported by embedding Class-I condition}

\author{Niyaz Uddin Molla \orcidlink{0000-0003-1409-2009}  }
\email{niyazuddin182@gmail.com}
\affiliation{Department of
Mathematics, Indian Institute of Engineering Science and Technology, Shibpur, Howrah-711 103, India}

\author{Himanshu Chaudhary}
\email{himanshuch1729@gmail.com}
\affiliation{Pacif Institute of Cosmology
and Selfology (PICS), Sagara, Sambalpur 768224, Odisha, India}

\author{Ujjal Debnath\orcidlink{ 0000-0002-2124-8908}}
\email{ujjaldebnath@gmail.com} \affiliation{Department of
Mathematics, Indian Institute of Engineering Science and
Technology, Shibpur, Howrah-711 103, India}

\author{G. Mustafa \orcidlink{0000-0003-1409-2009}}
\email{gmustafa3828@gmail.com}
\affiliation{Department of Physics,
Zhejiang Normal University, Jinhua 321004, People’s Republic of China}

\author{S. K. Maurya \orcidlink{0000-0003-4089-3651}}
\email{sunil@unizwa.edu.om} \affiliation{Department of Mathematical and Physical Sciences, College of Arts and Sciences, University of Nizwa, Nizwa 616, Sultanate of Oman}

\begin{abstract}
This study deals with the new class of embedded wormhole solutions in the background of general relativity. Two newly calculated wormhole solutions satisfy all the required properties. All the energy conditions are discussed through their validity regions for the different ranges of involved parameters. In maximum regions, all energy conditions are violated. We investigate the shadow and strong gravitational lensing by the wormhole throat for the two new wormhole models, namely Model-I and Model-II. The present paper considers the wormhole throat to act as a photon sphere. We first derive null geodesics using the Hamilton-Jacobi separation method to investigate the shadow and strong gravitational lensing caused by the wormhole throat. We then numerically obtain the radius of wormhole shadow, strong deflection angle, and various lensing observables by taking the example of supermassive black  M87* and  Sgr A* in the context of both Model-I and Model-II. Keeping all other parameters fixed, it is observed that the parameters $\zeta_1$ and $\zeta_2$ for Model-I; and  $\chi_1$ and $\chi_2$ for Model-II have significant effects on the wormhole shadow and strong gravitational lensing phenomena. Our conclusion is that it is possible to detect relativistic images, such as Einstein rings, produced by wormholes with throat radii of $r_{th}=3M$. Additionally, current technology enables us to test hypotheses related to astrophysical wormholes.
\end{abstract}

\maketitle

\date{\today}

\section{Introduction}  \label{sec1}
Wormholes, theoretical constructs speculated to exist within spacetime, have been a source of intrigue and investigation in the fields of theoretical physics and astrophysics. These traversable routes through spacetime, largely theoretical, still intrigue the scientific community because of their potential significance in comprehending the fundamental laws of the universe, particularly within the context of Einstein's theory of general relativity. The theoretical framework for wormholes was first developed by Flamm in 1916 \cite{flamm1916comments}. Later on, a comprehensive exploration of wormholes was undertaken by Albert Einstein and Nathan Rosen in their seminal paper in 1935 \cite{Einstein:1935tc} , commonly known as the Einstein-Rosen bridge. In the 20th century, the exploration of black holes and wormholes intersected as physicists delved deeper into the mathematical and physical principles governing these extraordinary phenomena. One of the most notable papers that advanced our comprehension of wormholes was published in 1988. This paper delved into the theoretical aspects of traversable wormholes and their possible applications in interstellar travel \cite{Morris:1988cz}. This research propelled wormholes into the forefront of astrophysical inquiry and introduced the notion of "exotic matter" as a crucial element for stabilizing these theoretical pathways through spacetime. In the literature, numerous authors have extensively examined various aspects of traversable wormhole (TW) geometries, as evidenced in References\cite{almheiri2020replica,Bronnikov:2002rn,penington2022replica,Hawking:1988ae,Lemos:2003jb,Visser:1989kh}.\\
Among the various intriguing aspects of wormholes, the investigation of their shadows and gravitational lensing phenomena offers a distinct opportunity to explore their properties and implications. In astrophysics, the concept of a shadow denotes the apparent outline of an object as perceived from a distant observer's viewpoint. This phenomenon is closely associated with gravitational lensing due to the bending of light around massive celestial bodies. Although shadows have been extensively studied with black holes, as evidenced in some literature \cite{Arora:2023ijd,Afrin:2023uzo,Ghosh:2023hnf,Ghosh:2022kit,Molla:2022izk} and therein, the investigation of shadows in wormholes represents a growing field with distinctive challenges and opportunities. The shadows created by wormholes offer valuable clues about their dimensions, structure, and the distribution of exotic matter inside them. This exploration can offer insights into the possibility of traversable wormholes and their observable manifestations.\\
Gravitational lensing serves as a highly valuable tool in astrophysics and cosmology for various purposes such as determining the cosmological constant, understanding mass distribution in the large-scale structure of the universe, mapping the distribution of dark matter, measuring the Hubble constant, studying galaxy cluster halos, and even confirming the existence of extrasolar planets (see details in Schneider et al.\cite{schneider1992gravitational}  and Perlick \cite{Perlick:2004tq,Perlick:2010zh}). At first, research on gravitational lensing primarily focused on theoretical exploration within the weak gravitational field. From the past few decades, there has been significant interest in exploring gravitational lenses under weak-field approximations \cite{petters2012singularity,Refsdal:1964yk,Refsdal:1964nw,Refsdal:1964yk,Refsdal:1964nw,Schmidt:2008hc,Jusufi:2017hed,Kumaran:2019qqp,Javed:2020frq}. Furthermore, research into gravitational lenses in strong gravitational fields have also been conducted, as these observations offer valuable insights into compact objects such as black holes. Darwin pioneered the study of gravitational lensing in a strong gravitational field \cite{darwin1959gravity}. The gravitational lensing effects of light rays, which are emitted by a source and pass around a compact lens object multiple times on a light sphere in a strong gravitational field, have been revisited several times \cite{atkinson1965light,Luminet:1979nyg,ohanian1987black,Virbhadra:1999nm,Bozza:2002zj,Nemiroff:1993he,Hasse:2001by,Bozza:2007gt,Bozza:2008ev,Bozza:2010xqn,Shaikh:2019itn,Kumar:2020sag,Hsieh:2021rru}
Wormholes also induce gravitational lensing effects in both weak and strong gravitational fields. The gravitational lenses created by wormholes were initially explored by Kim and Cho \cite{Kim:1994pe} and Cramer et al. \cite{Cramer:1994qj}. Subsequently, researchers investigated the gravitational lensing effects of negative mass wormholes \cite{Eiroa:2001zz,Safonova:2001vz,Safonova:2002si,takahashi2013observational} as well as positive mass wormholes \cite{Rahaman:2007am,Kuhfittig:2013hva,Kuhfittig:2013hva}. General relativity allows for nontrivial spacetime topology, such as wormhole spacetimes (refer to Visser \cite{visser1995lorentzian} for further information on wormholes). Several hypotheses regarding astrophysical wormholes have been explored \cite{Harko:2009xf,abdujabbarov2009electromagnetic}. For instance, Kardashev et al. propose that certain active galactic nuclei and other compact astrophysical objects could be explained as wormholes \cite{Kardashev:2006nj}. 
\par In this research paper, our objective is to explore shadows and gravitational lensing resulting from the presence of a wormhole throat. We focus on two distinct wormhole models to deepen our understanding. Utilizing mathematical formalism, we delve into the shadow and strong gravitational lensing phenomena induced by the wormhole, conceptualizing the wormhole throat as a photon sphere.
The structure of this study is planed as follows: Section~\ref{sec2} presents the Einstein field equations (EFEs). In Section~\ref{sec3}, we should calculate two new wormhole solutions by using embedded Class-I condition. Section~\ref{sec4} delves into discussions related to energy conditions with the supportive region under the effect of different involved parameters. In Section~\ref{sec5}, we discuss the  wormhole shadow by the wormhole throat for the two new wormhole models, namely Model-I and Model-II. Furthermore , strong gravitational lensing and its various observables for both the wormhole models should be discussed in Section \ref{sec6}. Finally, in Section~\ref{sec7} we have provided a few concluding remarks.

\section{Einstein field equations in the framework of modified matter source}\label{sec2}

The spherically symmetric static spacetime confined by a spherical surface $\Sigma$ is discussed in this section. With the scope of Schwarzschild coordinates, the line element is expressed as:
\begin{eqnarray}
ds^2 = e^{a(r)}dt^{2} - e^{b(r)}dr^{2}-r^{2}(\sin^{2}\theta d \phi^{2}+d\theta^{2}), \label{1}
\end{eqnarray}
where $b(r)$ and $a(r)$ are the candidates for the gravitational functions. The Einstein field equations through the matter sector for the metric (\ref{1}) can be written as:
\begin{eqnarray}
  G_{ij} = R_{ij}-\frac{1}{2} g_{ij} R = T_{ij}, \label{2}
\end{eqnarray}
We consider anisotropic matter distribution to investigate the system's internal structure with the matter source $T_{ij}$. Moreover, the anisotropic $T_{ij}$ energy-momentum tensor is given as
\begin{eqnarray}
T_{ij}=\rho\,u^i\,u_j-p\,K_{ij}+\Pi_{ij}, \label{eq4}
\end{eqnarray}
where
\begin{eqnarray}
&& p=\frac{2p_t + p_r}{3};~~~~\Pi_{ij}=\Pi \big(\xi^i \xi_j+\frac{1}{3} K_{ij}\big);\\
&& \Pi=p_r-p_t; ~~~K_{ij}=g_{ij}-u^i u_j, \label{eq5}
\end{eqnarray}
and the four-velocity vector $u^i$ and $\xi^i$ are given by
\begin{eqnarray}
u^i=(e^{-a/2},~0,~0,~0)~~\text{and}~~\xi^i=(0,~e^{-b/2},~0,~0), \label{eq6}
\end{eqnarray}
such that $\xi^i\xi_i=-1$ and $\xi^i u_i=0$. Finally, we present the following expressions for the matter source $T_{ij}$:
\begin{eqnarray}
&& T_{tt}=\rho,~~~T_{rr}=-p_r,~~~~T_{\theta\theta}=T_{\phi\phi}=-p_t, \label{eq7}
\end{eqnarray}
and the Einstein field equations (\ref{2}) read as
\begin{eqnarray}
\label{eq8}
&& \hspace{-0.5cm}\rho=\frac{1}{8\pi}\bigg[\frac{1}{r^2}-e^{-b}\left(\frac{1}{r^2}-\frac{b^{\prime}}{r}\right)\bigg],
\\\label{eq9}
&& \hspace{-0.5cm}{p}_{r}=\frac{1}{8\pi}\bigg[e^{-b}\left(\frac{1}{r^2}+\frac{a^{\prime}}{r}\right)-\frac{1}{r^2}\bigg],~~~~~ \\\label{eq10}
&& \hspace{-0.5cm}{p}_{t}=\frac{1}{8\pi}\bigg[\frac{e^{-b}}{4}\Big(2a^{\prime\prime}+a^{\prime2}-b^{\prime}a^{\prime}-2\frac{b^{\prime}-a^{\prime}}{r}\Big)\bigg],~~~
\end{eqnarray}
where primes signify derivatives with respect to $r$. Employing Schwarzschild coordinates according to the Morris and Thorne metric for  WH geometry, i.e., $e^{a(r)}=e^{2\Phi(r)}$ and $e^{b(r)}=\left(1-\frac{S(r)}{r}\right)^{-1}$~\cite{Morris:1988cz}, within the static spherically symmetric spacetime described by Equation (\ref{1}), we obtain the following system of field equations for WH geometry:
\begin{eqnarray}
\label{19}
&&\rho =\frac{S'(r)}{r^2},\\\label{20}
 && {p}_{r}=\frac{2  (r-S(r))r \Phi '(r)-S(r)}{r^3},~~~\\
\label{21}
 &&{p}_{t}= \frac{1}{4 r^3} \Big[\left(2 r \Phi '(r)+2\right)  \big(2 r^2 \Phi '(r)-2 r  \Phi '(r) S(r) -r S'(r) \nonumber\\&& \hspace{1.5cm} +S(r)\big) +4 r^2 (r-S(r)) \Phi ''(r)\Big].
\end{eqnarray}
In the above equations, $\Phi(r)$ and $S(r)$ are the representatives of the red-shift and shape function for the WH geometry. 
\begin{figure*}
\centering \epsfig{file=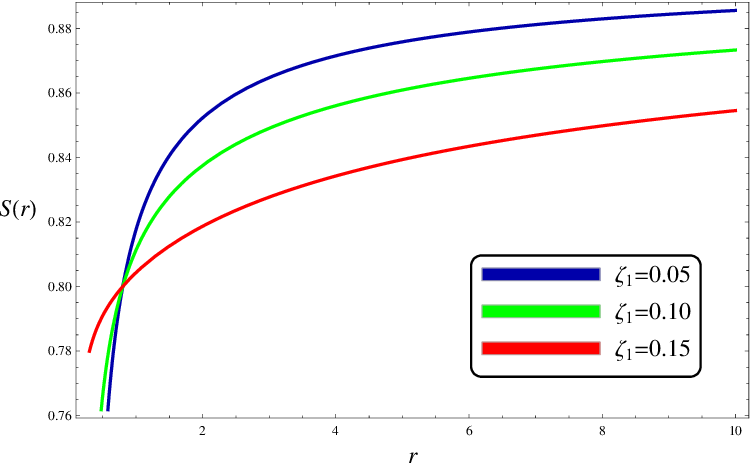, width=.3\linewidth,
height=2.5in} \epsfig{file=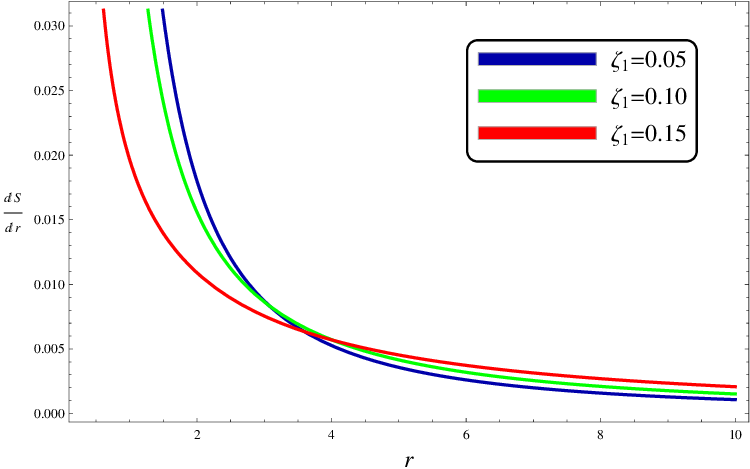, width=.3\linewidth,
height=2.5in} \epsfig{file=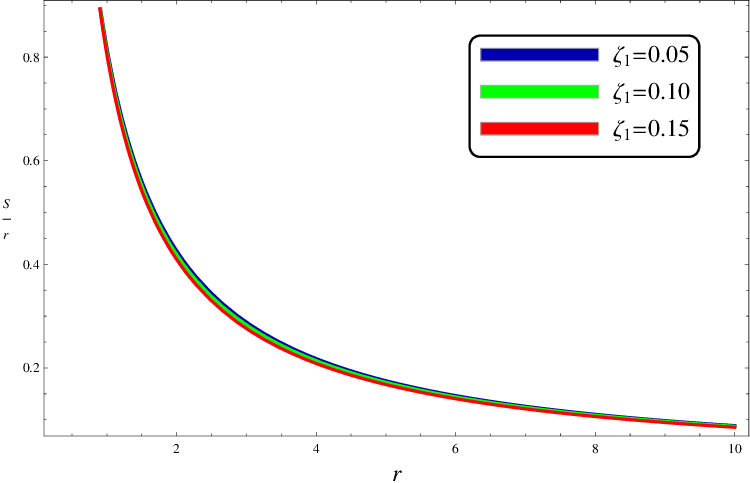, width=.3\linewidth,
height=2.5in}
\caption{\label{F1} shows the wormhole properties, including flatness and flaring out condition for model-I with three different values of $\zeta _1$.}
\end{figure*}
\begin{figure*}
\centering \epsfig{file=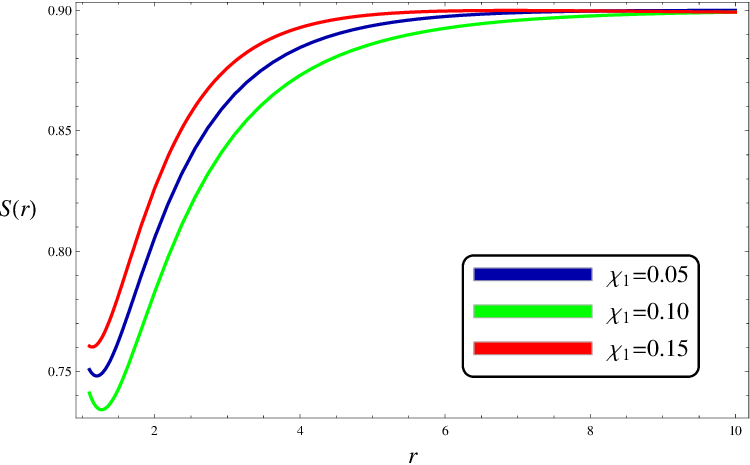, width=.3\linewidth,
height=2.5in} \epsfig{file=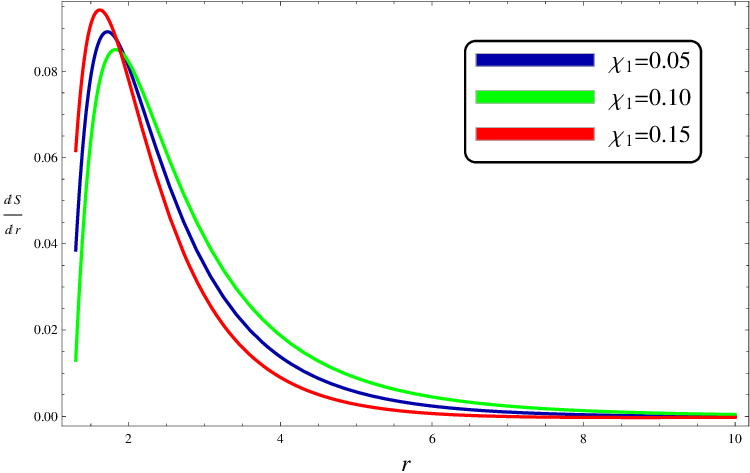, width=.3\linewidth,
height=2.5in} \epsfig{file=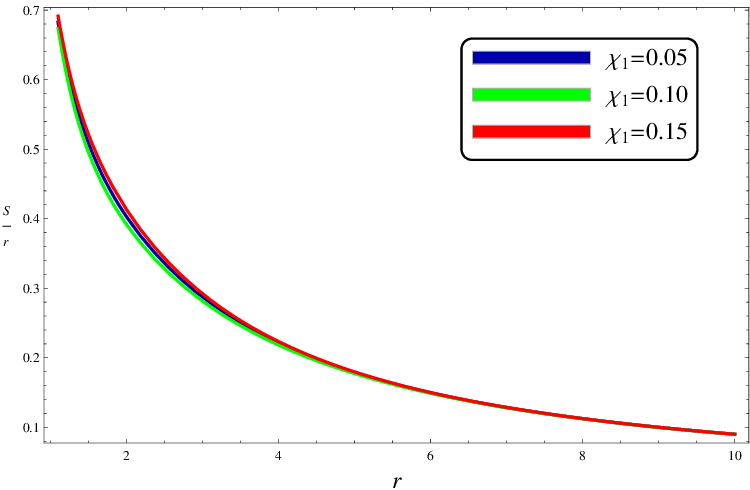, width=.3\linewidth,
height=2.5in}
\caption{\label{F2} shows the wormhole properties, including flatness and flaring out condition for model-II with three different values of $\chi _1$.}
\end{figure*}

\section{Embedding Class I condition and new wormhole solutions}\label{sec3}
In this study, we focus on utilizing embedded Class-I spacetime within the framework of the Karmarkar condition \cite{karmarkar1948gravitational} to explore a novel class of wormhole solutions. A pivotal component of our analysis revolves around the Karmarkar condition, which plays a crucial role in this investigation. Embedded Class-I solutions of Riemannian space serve as an essential prerequisite for establishing the foundation of the Karmarkar condition. Eisenhart \cite{eisenhart1997riemannian} proposed a vital condition for embedded Class-I solutions, which relies on two key elements: the Riemann curvature tensor denoted as \( R_{ab\alpha\beta} \), and a second-order symmetric tensor denoted as \( B_{ab} \). For a spacetime to be classified as a Class-I spacetime, whether it is spherically symmetric and static or non-static, it must fulfil specific necessary and sufficient conditions. These conditions encompass various aspects, and to be more explicit, any spherically symmetric spacetime, regardless of its static or non-static nature, needs to satisfy the following set of necessary and sufficient conditions. Firstly, a system featuring a symmetric tensor of the second order \( B_{ab} \) must establish a crucial relationship known as the Gauss equation, depicted as:
\begin{equation}
 R_{ab\alpha\beta}=\epsilon\left(B_{a\alpha} B_{b\beta}-B_{a\beta} B_{b\alpha}\right)   
\end{equation}
where \(\epsilon\) denotes whether the manifold is space-like (\(\epsilon=+1\)) or time-like (\(\epsilon=-1\)). Additionally, the symmetric tensor \( B_{ab} \) must satisfy Codazzi's equation, expressed as:
\begin{equation}
\nabla_{\alpha} B_{ab}-\nabla_{b} B_{a\alpha}=0
\end{equation}
These conditions form the backbone of our exploration and serve as pivotal guidelines in our pursuit of deriving novel wormhole solutions within the embedded Class-I spacetime framework and the constraints of the Karmarkar condition. In Schwarzschild's coordinates \( (t, r, \theta, \phi) \), where \( r = 0,1,2,3 \), the Riemannian components are computed as follows: 
\begin{equation}
\begin{aligned}
R_{r t r t} & = -e^{a}\left(\frac{a^{\prime \prime}}{2}-\frac{b^{\prime} a^{\prime}}{4}+\frac{a^{\prime 2}}{4}\right) ; \quad R_{r \theta r \theta} = -\frac{r}{2} b^{\prime} ; \\
R_{\theta \phi \theta \phi} & = -\frac{r^{2} \sin ^{2} \theta}{e^{b}}\left(e^{b}-1\right) ; \quad R_{r \phi \phi t} = 0 ; \\
R_{\phi t \phi t} & = -\frac{r}{2} a^{\prime} e^{a-b} \sin ^{2} \theta ; \quad R_{r \theta \theta t} = 0 .
\end{aligned}
\end{equation}
The Gauss equation can now be expressed as:
\begin{equation}
\begin{aligned}
& b_{t r} b_{\phi \phi} = R_{r \phi t \phi}=0 ; \quad b_{t r} b_{\theta \theta}=R_{r \theta t \theta}=0, \\
& b_{t t} b_{\phi \phi}=R_{t \phi t \phi} ; \quad b_{t t} b_{\theta \theta} = R_{t \theta t \theta} ; \quad b_{r r} b_{\phi \phi}=R_{r \phi r \phi},\\
& b_{\theta \theta} b_{\phi \phi} = R_{\theta \phi \theta \phi} ; \quad b_{r r} b_{\theta \theta}=R_{r \theta r \theta} ; \quad b_{t t} b_{r r}=R_{t r t r}.
\end{aligned}
\end{equation}
Utilizing the preceding relations, we derive the following expressions:
\begin{equation}
\begin{aligned}
& \left(b_{t t}\right)^{2}=\frac{\left(R_{t \theta t \theta}\right)^{2}}{R_{\theta \phi \theta \phi}} \sin ^{2} \theta, \quad\left(b_{r r}\right)^{2}=\frac{\left(R_{r \theta r \theta}\right)^{2}}{R_{\theta \phi \theta \phi}} \sin ^{2} \theta,\\
& \left(b_{\theta \theta}\right)^{2}=\frac{R_{\theta \phi \theta \phi}}{\sin ^{2} \theta}, \quad\left(b_{\phi \phi}\right)^{2}=\sin ^{2} \theta R_{\theta \phi \theta \phi}.
\end{aligned}\label{Rie4}
\end{equation}
An important connection in the Riemann components can be derived from Eq. ( \ref{Rie4})
\begin{equation}
R_{t \theta t \theta} R_{r \phi r \phi}=R_{t r t r} R_{\theta \phi \theta \phi}.
\end{equation}
This connection highlights the interrelationship among the various components of the Riemann tensor in the framework established by Eq.( \ref{Rie4}).
It's important to emphasize that Codazzi's equation arises from the equations presented above. A significant observation lies in the expression for the symmetric tensor $B_{ab}$ within the context of a non-static spherically symmetric spacetime. This tensor is governed by the following equation:
\begin{eqnarray}
    b_{t r} b_{\theta \theta}=R_{r \theta t \theta} \quad \text{and} \quad b_{t t} b_{r r}-\left(b_{t r}\right)^{2}=R_{t r t r}.
\end{eqnarray}
In the given equation, there exists a relationship expressed as $\left(b_{t r}\right)^{2}=\sin ^{2} \theta\left(R_{r \theta t \theta}\right)^{2} / R_{\theta \phi \theta \phi}$. In this context, the condition for the embedding Class-I can be expressed in the following manner:
\begin{eqnarray}
    R_{t \theta t \theta} R_{r \phi r \phi}=R_{t r t r} R_{\theta \phi \theta \phi}+R_{r \theta t \theta} R_{r \phi t \phi}.
\end{eqnarray}
The equation mentioned above is commonly referred to as the Karmarkar condition. However, this condition can be likened to the static spherically symmetric metric in our specific scenario. It serves as a crucial criterion for categorizing a spacetime as Class-I, offering both necessary and sufficient conditions for such classification. When we substitute the Riemann components with the aforementioned condition, the following differential equation is obtained:
\begin{eqnarray}
    2 \frac{a^{\prime \prime}}{a^{\prime}}+a^{\prime}=\frac{b^{\prime} e^{b}}{e^{b}-1}.
\end{eqnarray}
Upon integration of the given second-order differential equation, a relationship between gravitational potentials can be derived as follows:
\begin{eqnarray}\label{final}
    e^{b}=1+A a^{\prime 2} e^{a}, 
\end{eqnarray}
In the context where $A$ represents the integration constants, it becomes essential to consider a particular form of the red-shift function due to the existence of an embedded class one solution. This specific form of the red-shift function has been thoroughly explored.
\begin{equation}\label{red1}
 a(r)=2\Phi(r)=\frac{\zeta _2}{r^{\zeta _1}},   
\end{equation}
The constants $\zeta_1$ and $\zeta_2$ play crucial roles within our framework. We've chosen a red-shift function that satisfies the flatness condition. By following the methodology outlined in a previous study \cite{mustafa2021wormhole}, along with Eq. (\ref{red1}) and Eq. (\ref{final}), we arrive at the following relation for the shape function:
\begin{widetext}
\begin{equation}\label{shape1}
S(r)  = \frac{r (C_{1}-r_{0}) e^{\zeta _2 r^{-\zeta _1}} r_{0}^{2 \zeta _1+2}}{-C_{1} r^{2 \zeta _1+2} e^{\zeta _2 r_{0}^{-\zeta _1}}+C_{1} e^{\zeta _2 r^{-\zeta _1}} r_{0}^{2 \zeta _1+2}-e^{\zeta _2 r^{-\zeta _1}} r_{0}^{2 \zeta _1+3}}  +C_{1},\quad 0<C_{1}\leq r_{0}.
\end{equation} 
\end{widetext}
In our ongoing analysis, we'll refer to the shape function described by Eq. (\ref{shape1}) as Model-I. To broaden our investigation, we introduce another red shift function that meets the flatness condition. This function is defined as follows: 
\begin{equation}\label{red2}
 a(r)=2\Phi(r)=-2 r^{\chi _1} \omega ^{\chi _1}-\frac{2 \chi _2}{r}.   
\end{equation}
In Eq \ref{red2}, $\chi _1$ and $\chi _2$ represent constants. Following the approach outlined in \cite{mustafa2021wormhole}, alongside Eq. (\ref{red2}) and Eq. (\ref{final}), another generalized form of the embedded shape function can be derived as follows:
\begin{widetext}
\begin{eqnarray}\label{shape2}
S(r)=\frac{r r_{0}^4 S_1^2 (C_{2}-r_{0}) e^{2 r_{0}^{\chi _1} \omega ^{\chi _1}+\frac{2 \chi _2}{r_{0}}}}{C_{2} \left(r_{0}^4 S_1^2 e^{2 r_{0}^{\chi _1} \omega ^{\chi _1}+\frac{2 \chi _2}{r_{0}}}-r^4 S_1^2 e^{2 r^{\chi _1} \omega ^{\chi _1}+\frac{2 \chi _2}{r}}\right)-r_{0}^5 S_1^2 e^{2 r_{0}^{\chi _1} \omega ^{\chi _1}+\frac{2 \chi _2}{r_{0}}}}+C_{2},\;\;\; 0<C_{2}\leq r_{0},
\end{eqnarray}
\end{widetext}
where 
\begin{equation*}
S_1=\chi _2-\chi _1 r^{\chi _1+1} \omega ^{\chi _1}
\end{equation*}
\begin{figure*}
\centering \epsfig{file=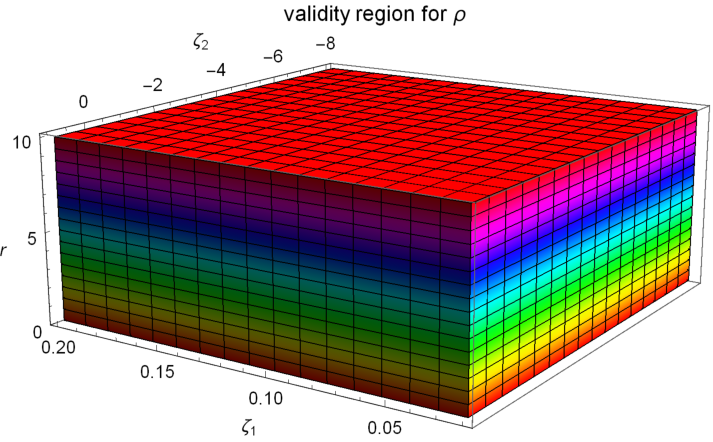, width=.48\linewidth,
height=2.5in} \epsfig{file=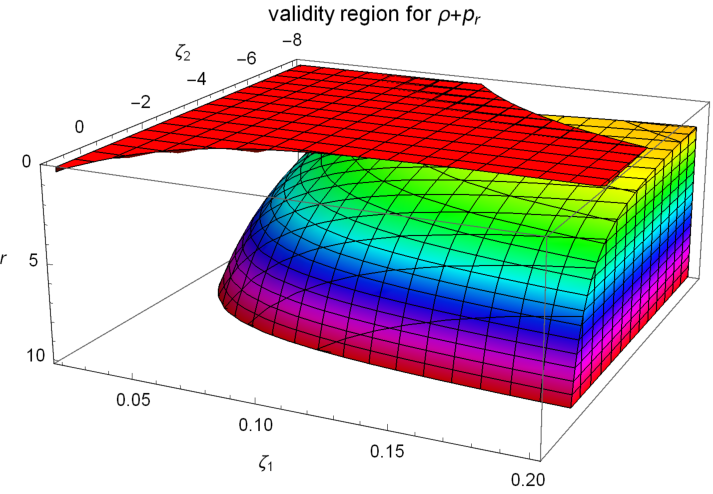, width=.48\linewidth,
height=2.5in}
\caption{\label{F3} shows the valid region of $\rho$ and $\rho+p_r$ for first shape function.}
\end{figure*}

\begin{figure*}
\centering \epsfig{file=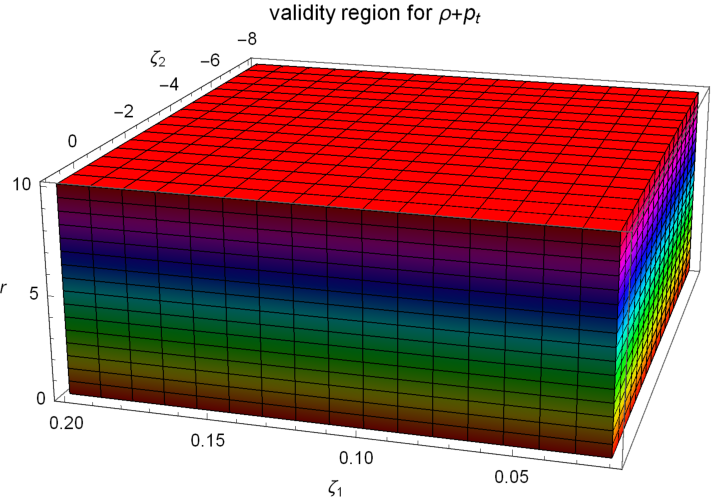, width=.48\linewidth,
height=2.5in} \epsfig{file=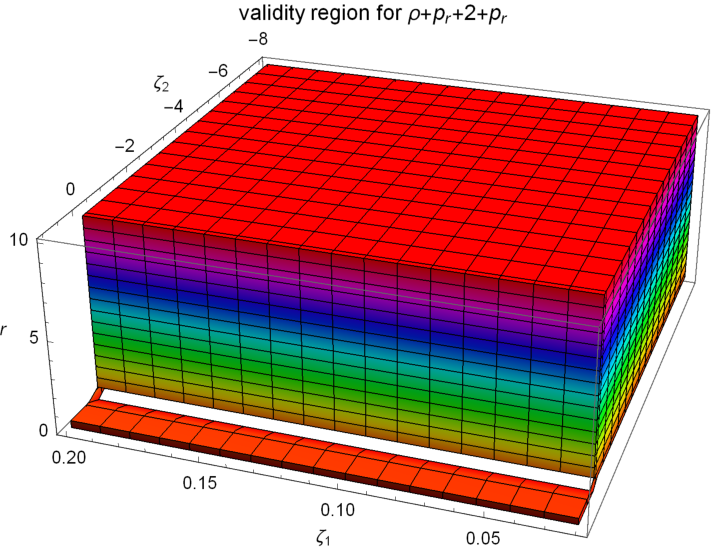, width=.48\linewidth,
height=2.5in}
\caption{\label{F4} shows the valid region of $\rho+p_t$ and $\rho+p_{r}+2p_t$ for first shape function.}
\end{figure*}

\begin{figure*}
\centering \epsfig{file=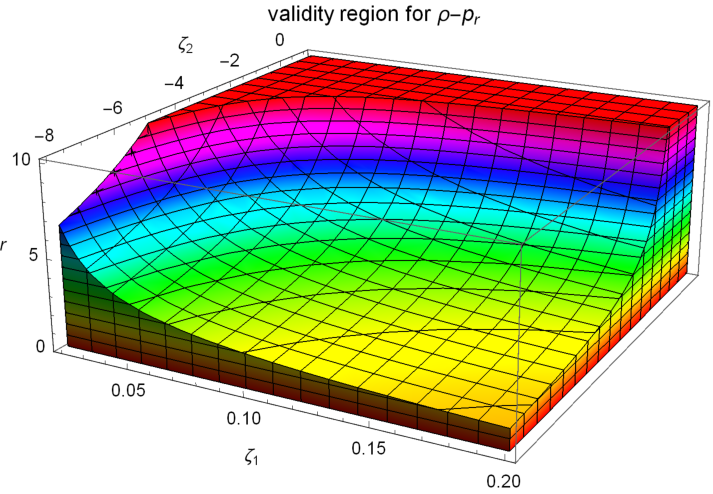, width=.48\linewidth,
height=2.5in} \epsfig{file=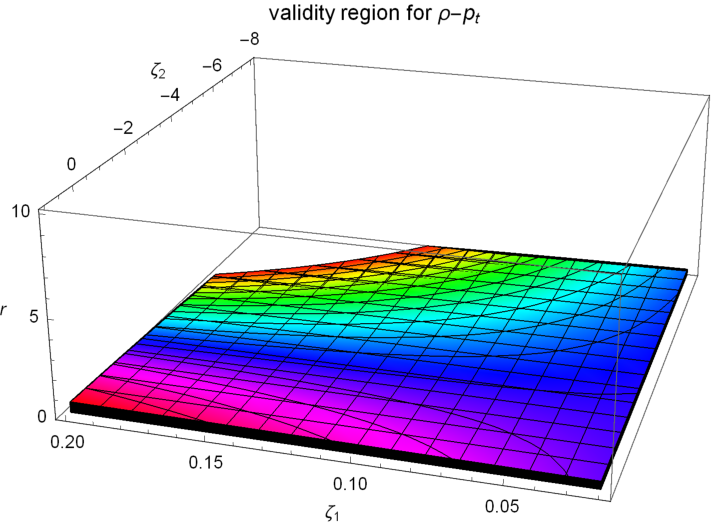, width=.48\linewidth,
height=2.5in}
\centering \epsfig{file=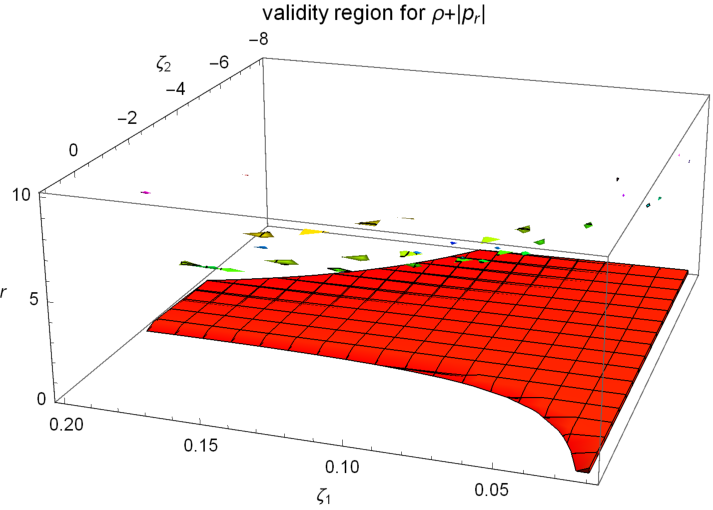, width=.48\linewidth,
height=2.5in} \epsfig{file=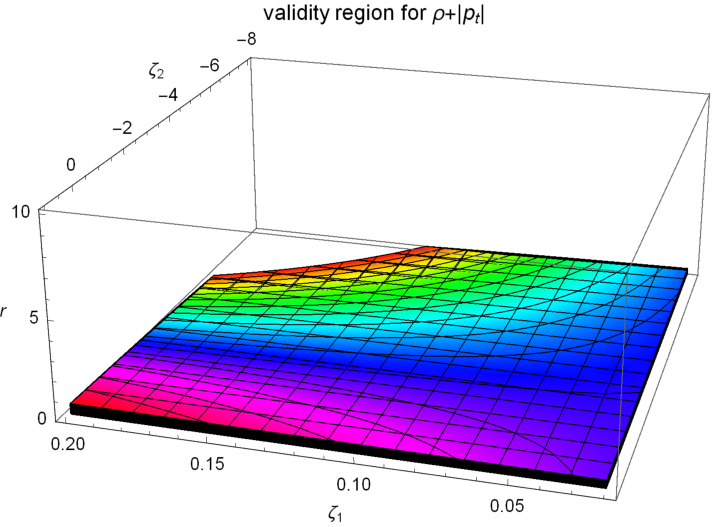, width=.48\linewidth,
height=2.5in}
\caption{\label{F5} shows the valid region of $\rho-p_r$ and $\rho-p_t$ is given in first row. The valid region of $\rho-|p_r|$ and $\rho-|p_t|$ is also presented in the second row for the first shape function.}
\end{figure*}

\begin{figure*}
\centering \epsfig{file=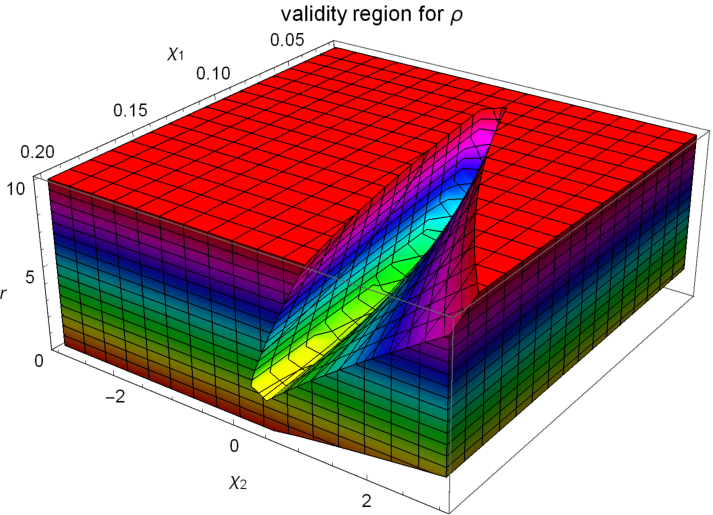, width=.48\linewidth,
height=2.5in} \epsfig{file=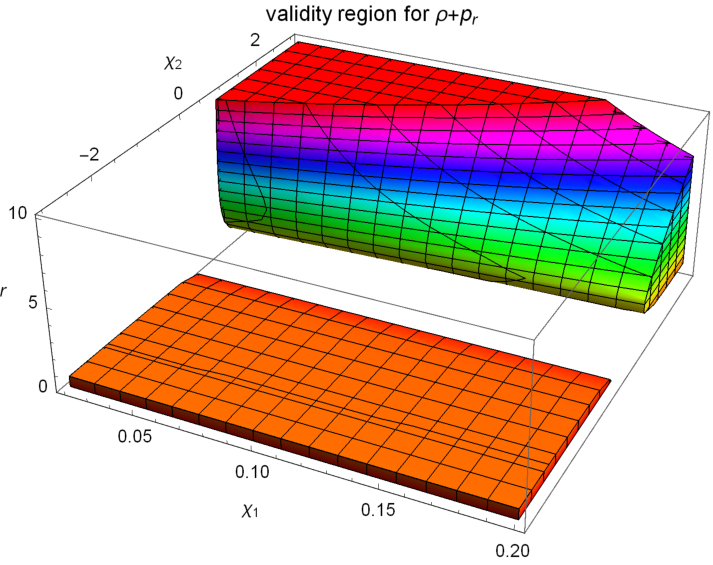, width=.48\linewidth,
height=2.5in}
\caption{\label{F6} shows the valid region of $\rho$ and $\rho+p_r$ for second shape function}
\end{figure*}

\begin{figure*}
\centering \epsfig{file=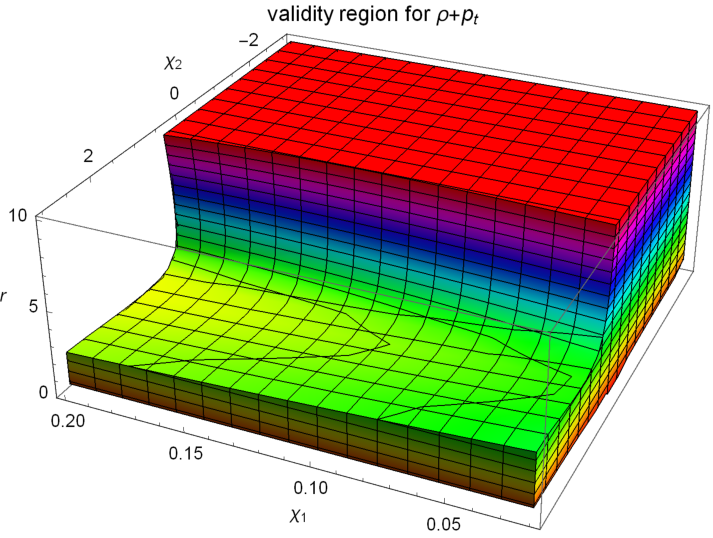, width=.48\linewidth,
height=2.5in} \epsfig{file=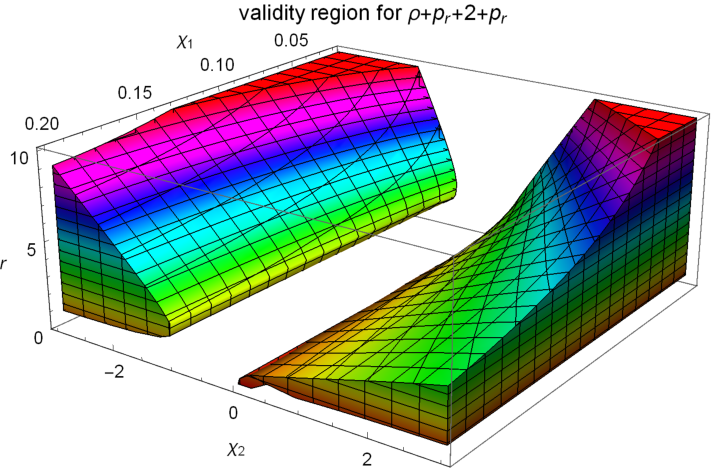, width=.48\linewidth,
height=2.5in}
\caption{\label{F7} shows the valid region of $\rho+p_t$ and $\rho+p_{r}+2p_t$ for first second function.}
\end{figure*}

\begin{figure*}
\centering \epsfig{file=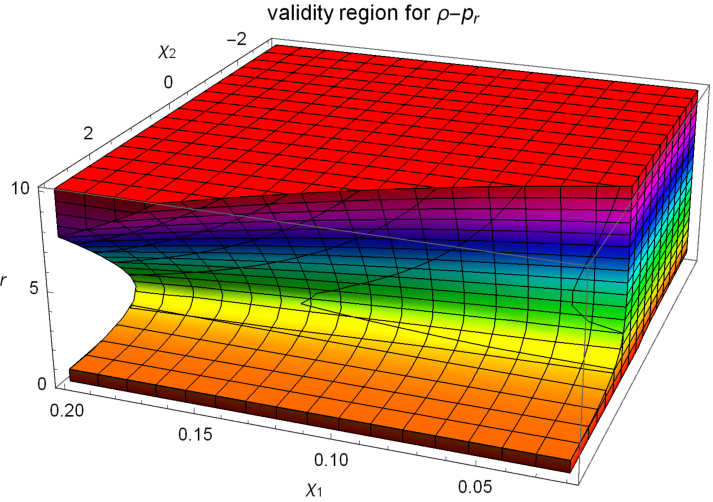, width=.48\linewidth,
height=2.5in} \epsfig{file=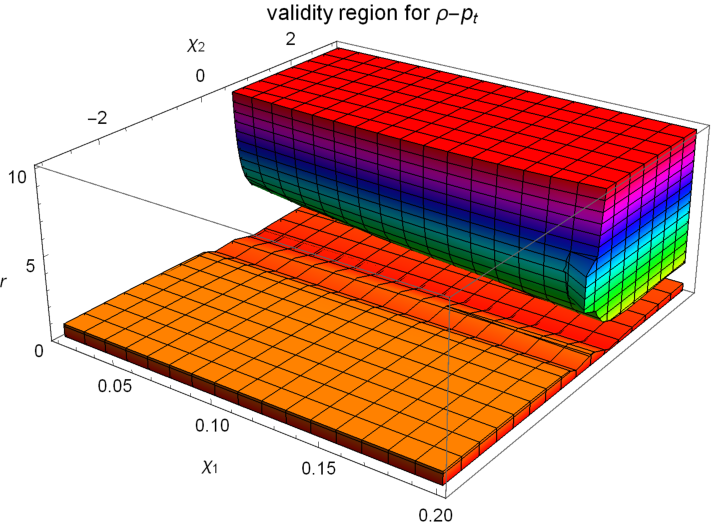, width=.48\linewidth,
height=2.5in}
\centering \epsfig{file=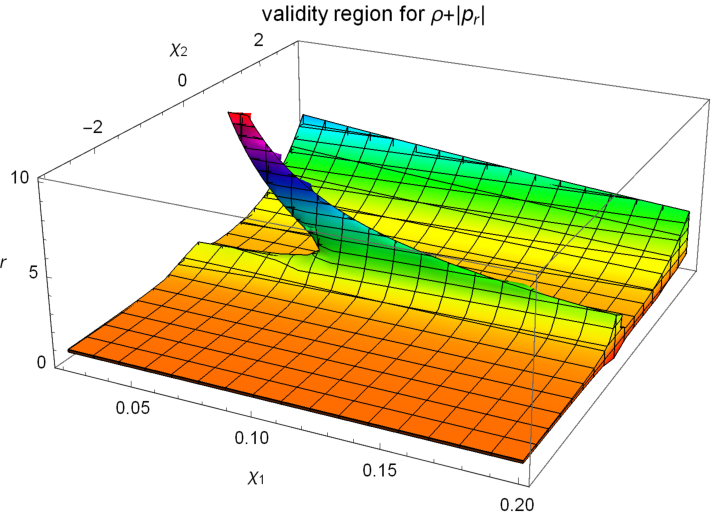, width=.48\linewidth,
height=2.5in} \epsfig{file=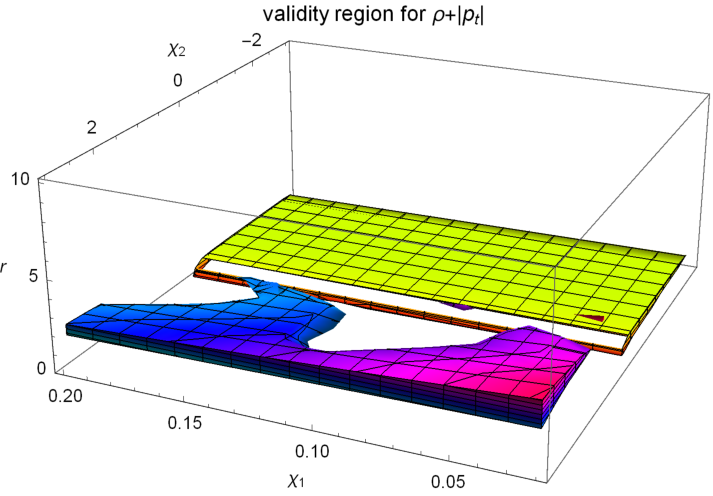, width=.48\linewidth,
height=2.5in}
\caption{\label{F8} shows the valid region of $\rho-p_r$ and $\rho-p_t$ is given in first row. The valid region of $\rho-|p_r|$ and $\rho-|p_t|$ is also presented in the second row for the second shape function.}
\end{figure*}
A wormhole's shape function explains the geometry of the passage through the wormhole and determines its shape. The shape function of a wormhole has the following characteristics: From the minimum throat radius, shape function should be increased with positive behavior. From Fig. (\ref{F1}) and Fig. (\ref{F2}) by the left part, it can be seen that both the calculated shape functions are continuously increasing with a positive trend throughout the described configurations. Flaring out and flatness conditions are also satisfied for the calculated shape functions. The graphical behavior from Fig. (\ref{F1}) and Fig. (\ref{F2}) by middle and right plots confirm the required behavior for flaring out and flatness conditions. The wormhole's shape function under the above-mentioned conditions could be traversable. It is necessary to mention that the flatness condition is a fundamental property that helps connect two space regions smoothly without any singularities.  

\section{Regional sustainability of Energy Conditions }\label{sec4}

In GR, energy conditions refer to constraints on the movement of matter and energy in spacetime that must be met for the theory to be valid. The primary energy bounds are the weak energy condition (WEC), null energy condition (NEC), dominant energy condition (DEC), and strong energy condition (SEC). The above conditions are described as
 ~WEC$\Leftrightarrow T_{ab}X^a X^b\geq 0,$ ~DEC:$\Leftrightarrow T_{ab}X^a X^b\geq 0$,
  NEC:$\Leftrightarrow T_{ab}\chi^a \chi^b\geq 0$, and SEC: $\Leftrightarrow (T_{ab}-\frac{T}{2}g_{ab})X^a X^b\geq 0$. 
with $\chi^a$ is the null vector and $X^a$ is a time-like vector. For $DEC$, $T_{ab}X^a$ is not space like. Further, for the principal pressure, all the energy conditions are discussed as DEC: $ \Leftrightarrow\rho\geq 0,~~\forall j,~p_j\in[-\rho,+\rho],$ SEC: $\Leftrightarrow \forall j (j=r,t),~p_j+\rho\geq 0,~\sum_jp_j+\rho\geq 0,$ WEC: $ \Leftrightarrow\rho\geq 0,~~\forall j,~p_j+\rho\geq 0,$ and NEC: $\Leftrightarrow\forall j,~p_j+\rho\geq 0,$~ Finally, we have following updated version of all the energy conditions: SEC:$p_{r}+\rho\geq 0,p_t+\rho\geq 0,p_r+2p_t+\rho\geq 0,$ WEC:$\rho\geq 0,p_r+\rho\geq 0,p_t+\rho\geq 0,$ NEC:$p_r+\rho\geq 0,p_t+\rho\geq 0,$ and
DEC:$\rho\geq 0,-|p_r|+\rho\geq 0,-|p_t|+\rho\geq 0.$
The corresponding evolution of radial and tangential pressures and energy density are the most essential conditions for the creation of wormhole structures. The main purpose of this study is to investigate the evolution of energy conditions under the anisotropic source of matter. Fig. (\ref{F3}) and Fig. (\ref{F6}) describe the regional representation of energy density and NEC, specifically the $\rho+p_r$ expression for both calculated shape functions, respectively. The violated behavior of energy conditions in the context of wormholes is needed to keep the wormhole throat open. All other energy conditions are provided in Fig. (\ref{F4}) and Fig. (\ref{F5}) for model-I. For the second model, all remaining energy conditions are presented by valid region in  Fig. (\ref{F7}) and Fig. (\ref{F8}). All the energy conditions, including SEC, are violated in the maximum region for the different ranges of involved parameters. The violated behavior, especially NEC and SEC, confirms the presence of unusual kinds of matter, say exotic matter. The presence of exotic matter is a necessary condition for the creation of wormholes. Exotic matter that has not yet been detected or shown to be physically feasible, such as warp wormholes, may occur if these energy conditions are violated. It is important to remember that these energy conditions are helpful resources for learning about the characteristics of wormhole geometry in GR. In GR, negative energy conditions are required to develop and sustain the exotic matter thought to be present in wormholes. This exotic matter shows characteristics not found in normal matter, such as negative mass and negative energy.




\section{ Shadows of Wormhole}\label{sec5} 
In this section, we discuss the  wormhole shadow using the static spherically symmetric wormhole spacetime \eqref{1}  by the wormhole throat for the two new wormhole models, namely Model-I and Model-II.
To investigate the light deflection influenced by the wormholes, we must quantify how the light ray moves around them. We consider the path of a light ray following the null geodesic equation, allowing us to predict the light trajectory.
Since the wormhole spacetime \eqref{1} is static and spherically symmetric, the projection of photon four-momentum along the killing vectors of isometries yields conserved quantities such as the energy $E=-p_{\mu}\xi^{\mu}_{(t)}$ and the angular momentum $L=-p_{\mu}\xi^{\mu}_{(\phi)}$, which remain constant along the geodesics. Here, $\xi^{\mu}{(t)}$ and $\xi^{\mu}_{(\phi)}$ represent the killing vectors associated with time translation and rotational invariance, respectively.

This leads to the following expressions on the equatorial plane $\frac{\pi}{2}$:

\begin{equation}\label{28}
\dot{t} = \frac{dt}{d\tau} = \frac{E}{e^{2\phi(r)}}
\end{equation}

\begin{equation}\label{29}
\dot{\phi} = \frac{d\phi}{d\tau} = \frac{L}{r^2 }
\end{equation}

where $\tau$ is the affine parameter.

Using the Eqs. (\ref{28}) and (\ref{29}), one can obtain the orbit equation as
\begin{equation}\label{30}
   K_{E}+V_{eff}=\frac{1}{\Tilde{b}^2}
\end{equation}
where  $\Tilde{b}=\frac{L}{E}$  the  kinetic energy function $K_{E}$ and   potential function $V_{eff}$ is described by
\begin{equation}\label{31}
 K_{E}=  \frac{e^{2\phi(r)}}{1-\frac{S(r)}{r}}\dot{r}^2
\end{equation}

\begin{equation}\label{32}
 V_{eff}=   \frac{e^{2\phi(r)}}{r^2}
\end{equation}
\par

\begin{figure*}
\centering
\includegraphics[width=.45\textwidth]{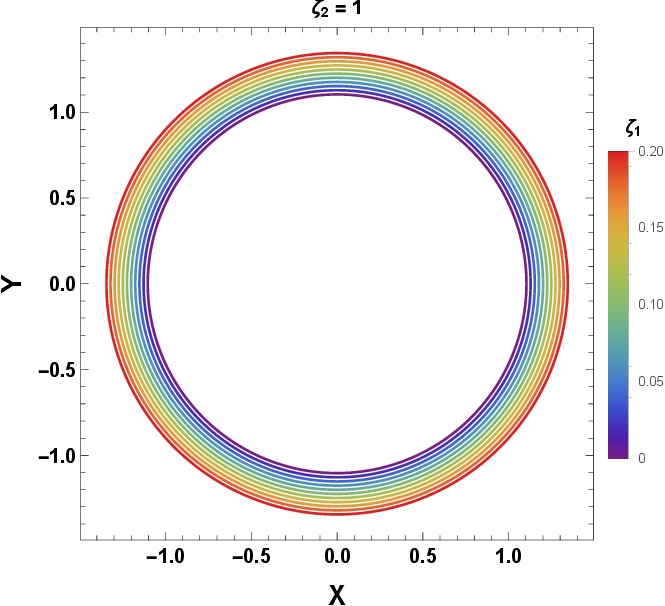}(a)
\qquad
\includegraphics[width=.45\textwidth]{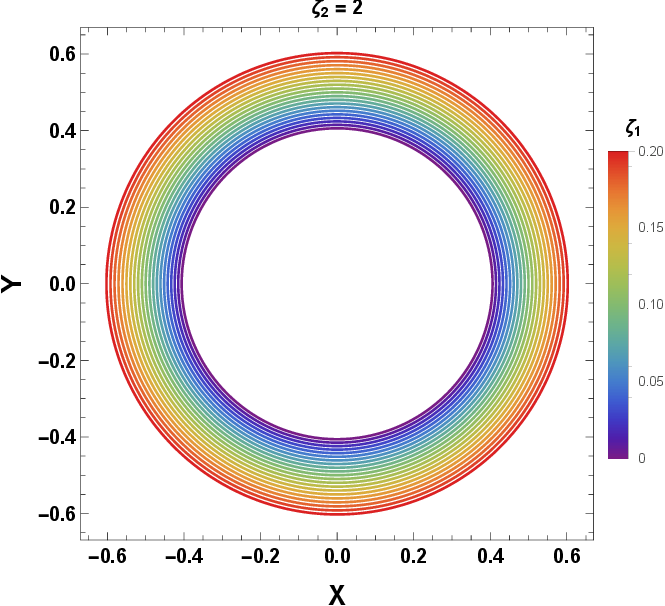}(b)
\caption{Shadow silhouette of the wormhole for $\zeta_2=1$ (left panel)  and $\zeta_2=2$ (right panel) with varying $\zeta_1$  as seen from the equatorial plane i.e. $\theta_0=\pi/2$ for wormhole space-time Model-I}\label{fig:9}
\end{figure*}

\begin{figure*}
\centering
\includegraphics[width=.45\textwidth]{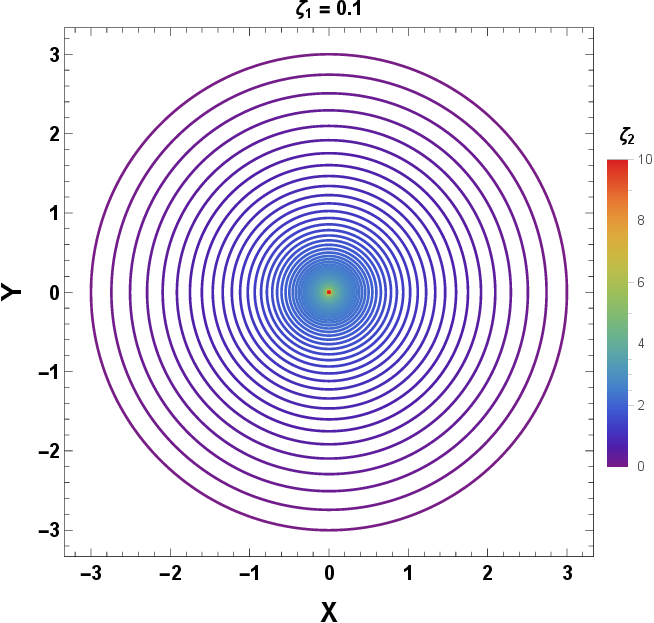}(a)
\qquad
\includegraphics[width=.45\textwidth]{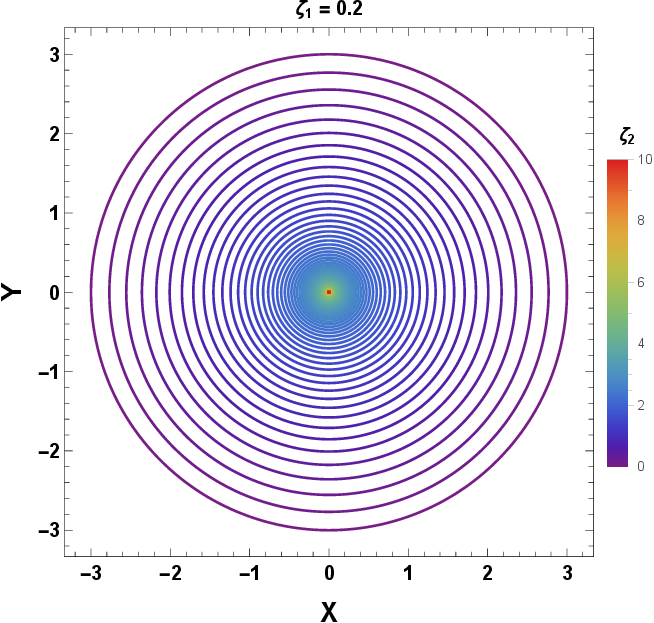}(b)
\caption{Shadow silhouette of the wormhole for $\zeta_1= 0.1$ (left panel)  and $\zeta_1= 0.2$ (right panel) with varying $\zeta_2$  as seen from the equatorial plane i.e. $\theta_0=\pi/2$ for wormhole space-time Model-I.}\label{fig:10}
\end{figure*}

\begin{figure*}
\centering
\includegraphics[width=.45\textwidth]{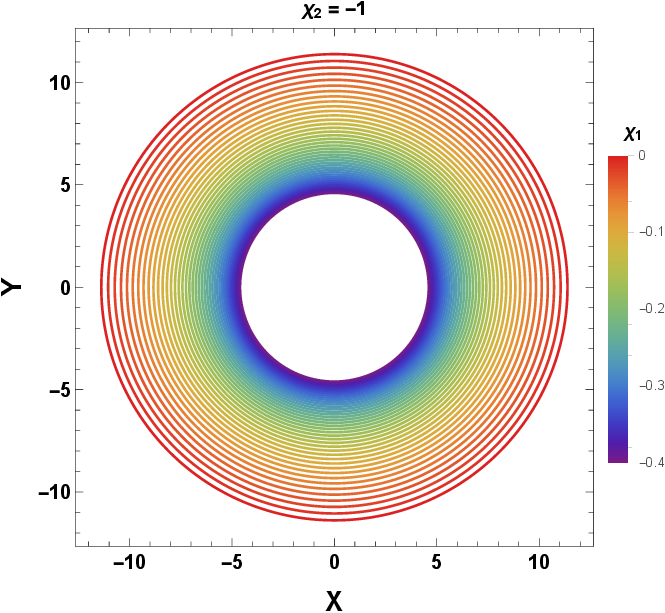}(a)
\qquad
\includegraphics[width=.45\textwidth]{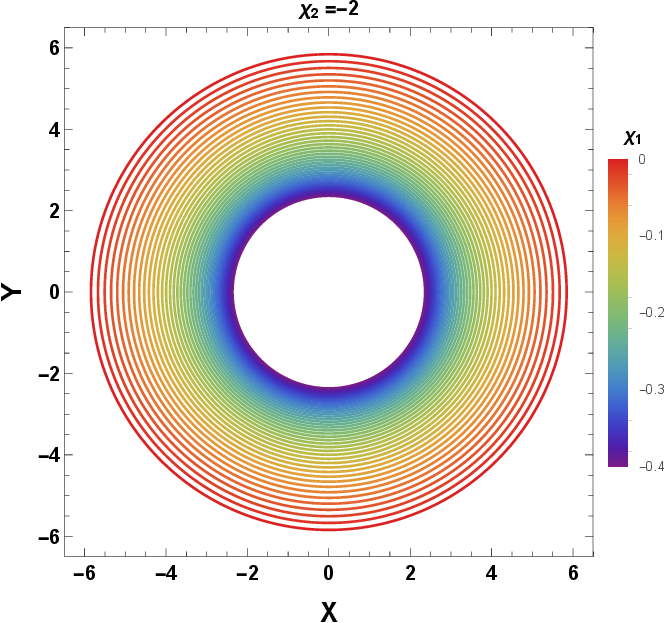}(b)
\caption{Shadow silhouette of the wormhole for  $\chi_2=-1,$ (left panel)  and  $\chi_2= -2$ (right panel) with varying $\chi_1$  as seen from the equatorial plane i.e. $\theta_0=\pi/2$ for wormhole space-time Model-II.
}\label{fig:11}
\end{figure*}

\begin{figure*}
\centering
\includegraphics[width=.45\textwidth]{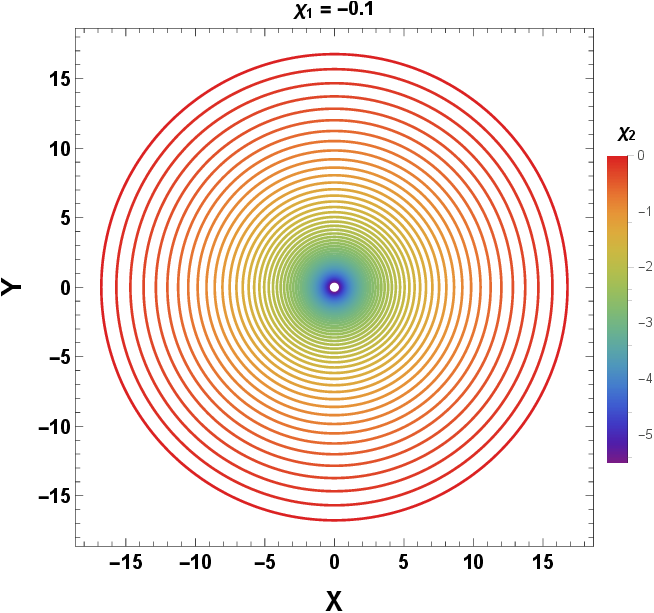}(a)
\qquad
\includegraphics[width=.45\textwidth]{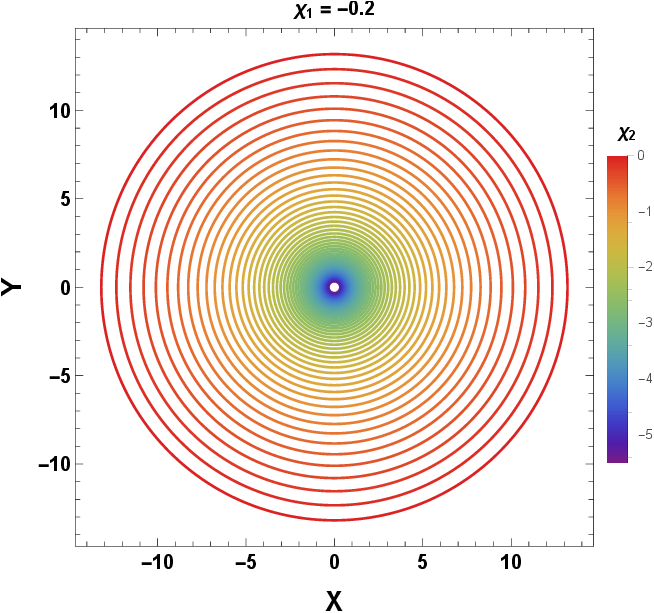}(b)
\caption{Shadow silhouette of the wormhole for  $\chi_1=-0.1,$ (left panel)  and  $\chi_1= -0.2$ (right panel) with varying $\chi_2$  as seen from the equatorial plane i.e. $\theta_0=\pi/2$ for wormhole space-time Model-II.}\label{fig:12}
\end{figure*}

To describe the wormhole shadow boundary, one can introduce the celestial coordinates ($X,Y$) defined as
\begin{equation}\label{33}
X=\lim_{r_0\rightarrow \infty}(-r_0^2  \sin(\theta_0))\frac{d\phi}{dr},
\quad
Y=\lim_{r_0\rightarrow \infty}(r_0^2  \frac{d\theta}{dr}),
\end{equation}

where $r_0$ is the radial distance of the wormhole with respect to the observer and $\theta_0$ is the inclination angle between the wormhole and the observer. 
Assuming the static observer is located at infinity, the radius of the wormhole shadow $r_s$ as seen from the equatorial plane, i.e. $\theta_0=\pi/2$ can be expressed as:
\begin{equation}\label{34}
r_s=\sqrt{X^2 +Y^2}=\frac{r_{th}}{e^{\phi(r_{th})}}
\end{equation}
The parametric plot for the Eq. (\ref{34}) in the $(X, Y)$ plane can cast a variety of wormhole shadows. The static spherically symmetric wormhole shadows for the wormhole spacetime have been depicted in Figs. (\ref{fig:9}), (\ref{fig:10}) for the wormhole Model-I and Figs. (\ref{fig:11}) and (\ref{fig:12}) for the wormhole Model-II. From Figs. (\ref{fig:9}) and  (\ref{fig:10}), it is observed that the radius of the wormhole shadow for the wormhole Model-I increases with $\zeta_1$ for a fixed value of $\zeta_2=1$ or $2$, while it decreases with $\zeta_2$ for a fixed value of $\zeta_2=0.1$ or $0.2$. However, from Figs. (\ref{fig:11}) and (\ref{fig:12}), it is observed that the radius of the wormhole shadow for the wormhole Model-II increases with $\chi_1$ for a fixed value of $\chi_2=-1$ or $-2$, and also increases with $\chi_2$ for a fixed value of $\chi_2=-0.1$ or $-0.2$.



\section{Strong Gravitational Lensing and Its Observable due to a wormhole  throat}\label{sec6} 
Here, we study the strong gravitational lensing due to a wormhole throat with two different wormhole models. Here, we investigate the deflection of photon rays in the equatorial plane ($\theta=\frac{\pi}{2}$) due to due to a wormhole throat, where the wormhole throat itself acts as an effective photon sphere, i.e., as
a position of the maximum of the effective potential for photons. To study the  deflection angle of photon rays in the equatorial plane ($\theta=\frac{\pi}{2}$), the  static, spherically symmetric wormhole metric for both Model-I and  Model-II can be expressed as:
 \begin{equation}\label{35}
d\bar{s}^2=-A(r)dt^2+ B(r) dr^2 +C(r) d\phi^2, 
\end{equation}
where  $A(r)=e^{2\Phi(r)}$ , $B(r)=\left(1-\frac{S(r)}{r}\right)^{-1}$ and  $C(r)=r^2$ for both the Model-I and  Model-II.
Note that the metric functions satisfy the asymptotically flat conditions for both  Model-I and  Model-II with viable parameter ranges.
\begin{equation}\label{36}
  \lim_{r\rightarrow \infty}A(r)=1,
 \lim_{r\rightarrow \infty}B(r)=1,
  \lim_{r\rightarrow \infty}C(r)=1
\end{equation}

  \begin{figure*}
\centering
\includegraphics[width=.45\textwidth]{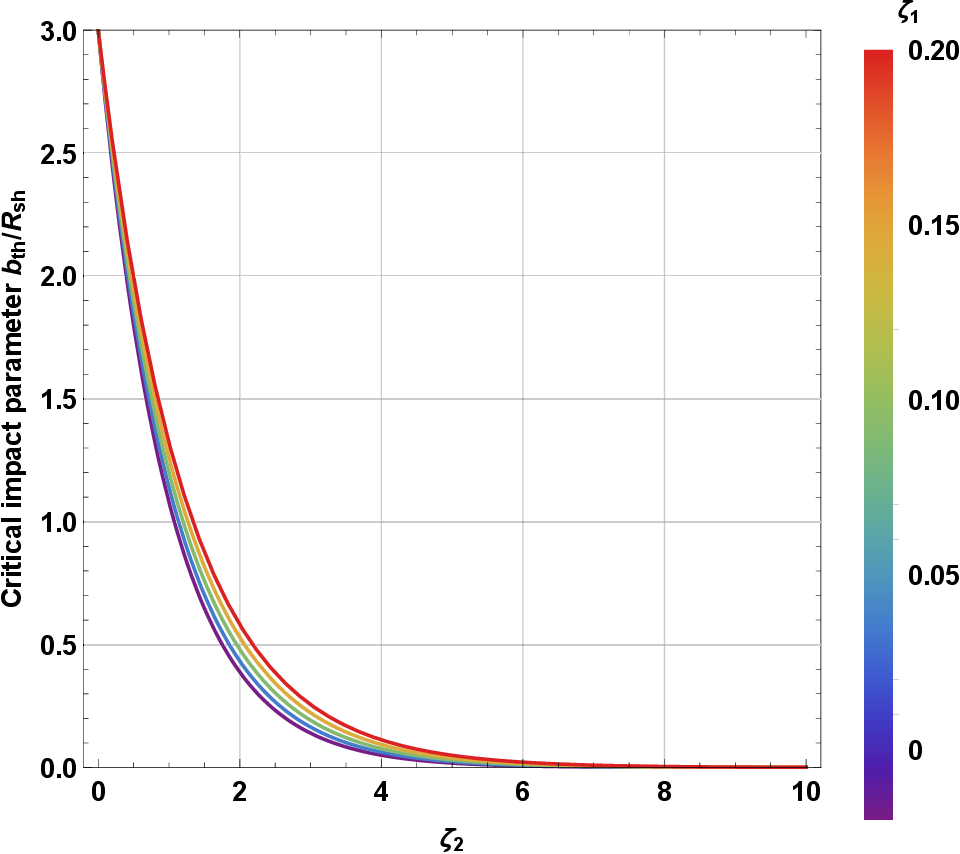}(a)
\qquad
\includegraphics[width=.45\textwidth]{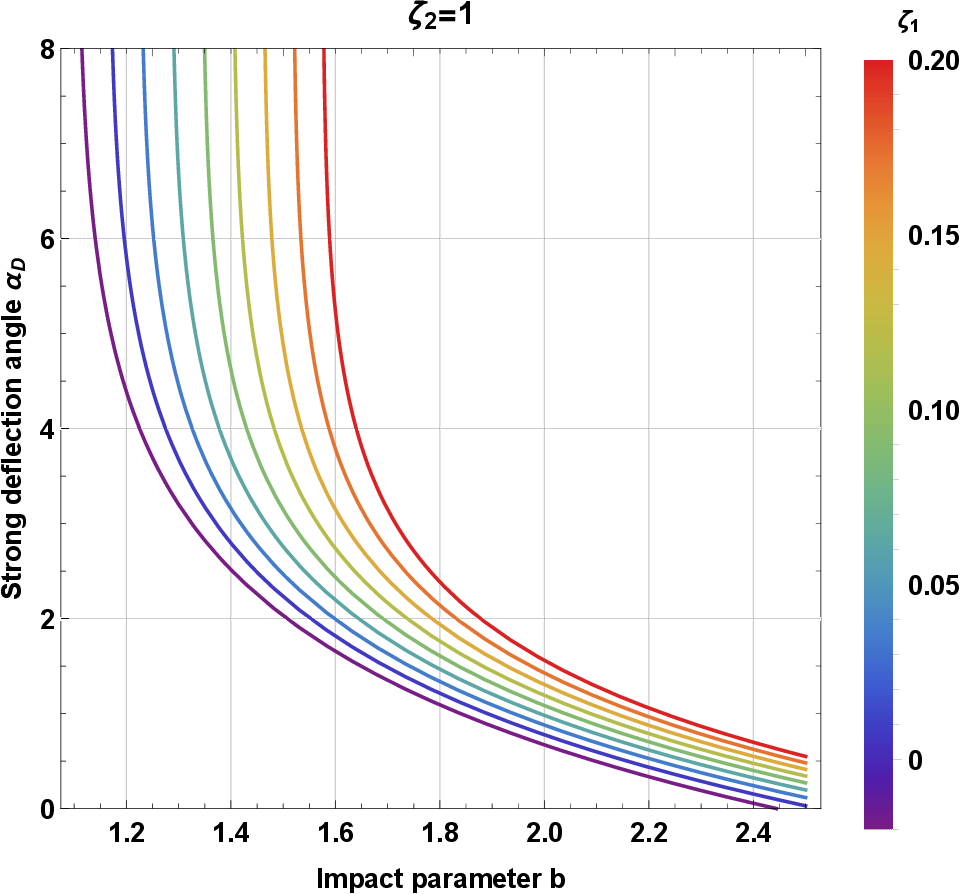}(b)
\caption{The behavior of the critical impact parameter $\mathit{u_{ph}/R_{sh}}$ (left panel) and strong deflection angle $\alpha_D $ (right panel) with  both the parameters $\zeta_1$ and $\zeta_2$ for the wormhole space-time Model-I..
}\label{fig:13}
\end{figure*}

\begin{figure*}
\centering
\includegraphics[width=.45\textwidth]{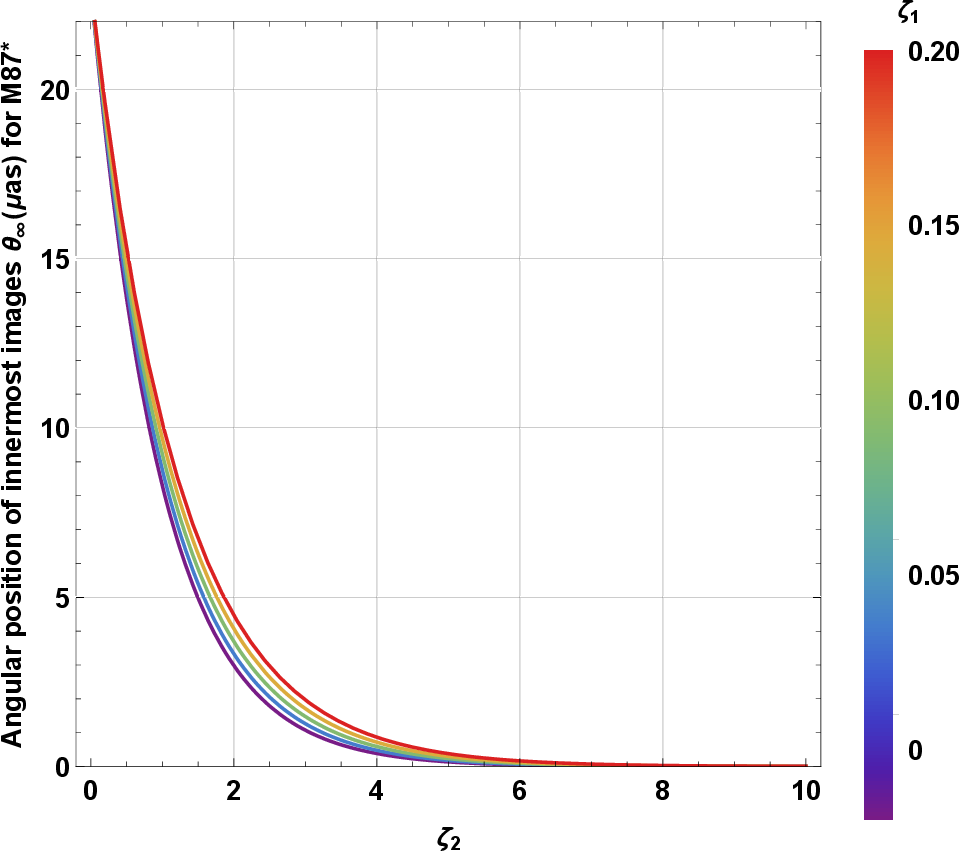}(a)
\qquad
\includegraphics[width=.45\textwidth]{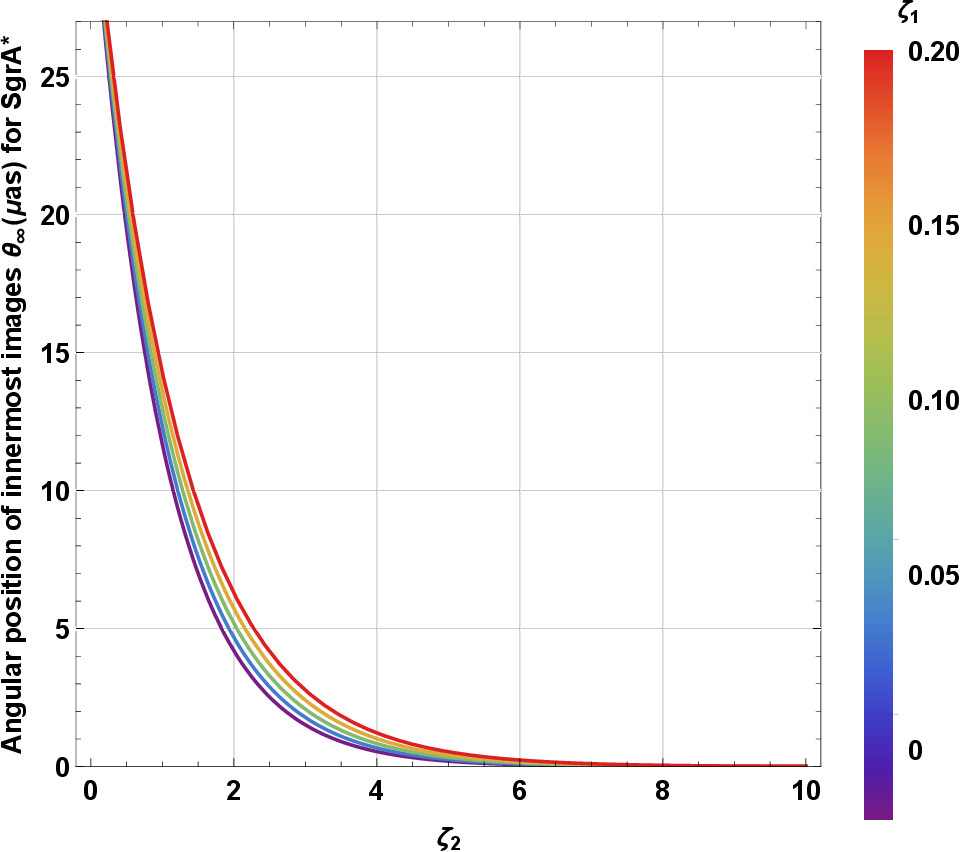}(b)
\caption{The behaviour of the angular position of the innermost image $\theta_{\infty}$ for M87* (left panel) and Sgr A* (right panel)   with  both the parameters $\zeta_1$ and $\zeta_2$ for the wormhole space-time Model-I.}\label{fig:14}
\end{figure*}
\begin{figure*}
\centering
\includegraphics[width=.45\textwidth]{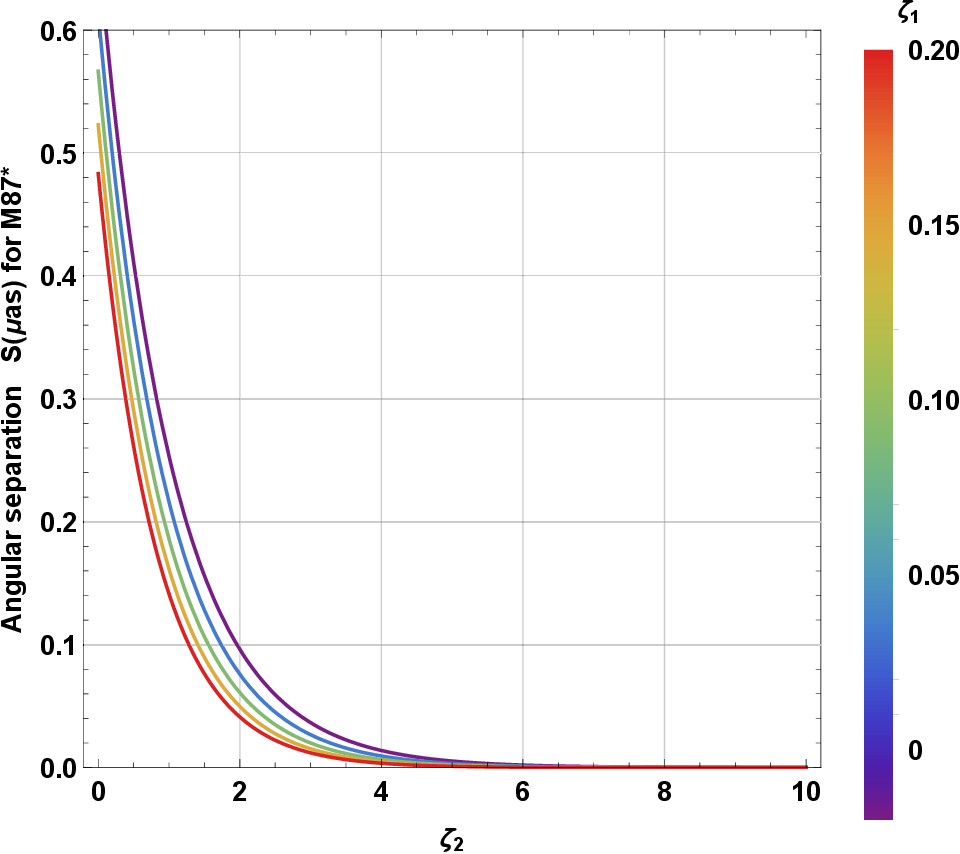}(a)
\qquad
\includegraphics[width=.45\textwidth]{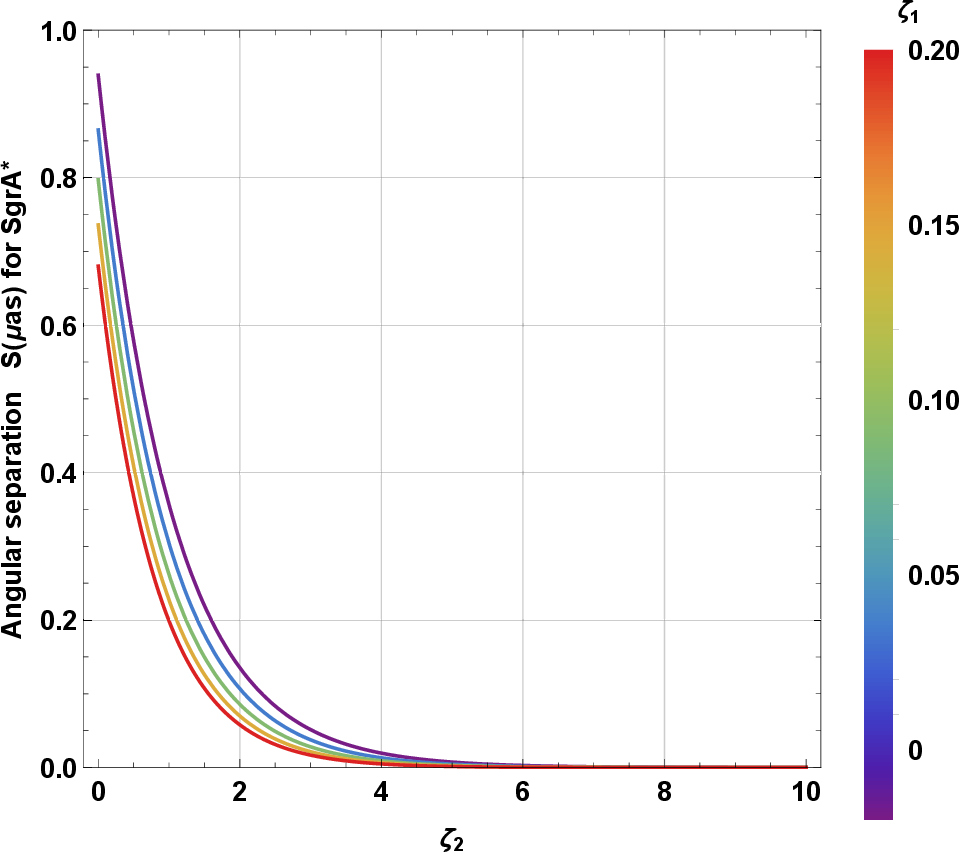}(b)
\caption{The behaviour of the angular separation of the innermost and outermost images $S$ for M87* (left panel) and Sgr A* (right panel)   with  both the parameters $\zeta_1$ and $\zeta_2$for the wormhole spacetime Model-I.}\label{fig:15}
\end{figure*}
\begin{figure*}
\centering
\includegraphics[width=.45\textwidth]{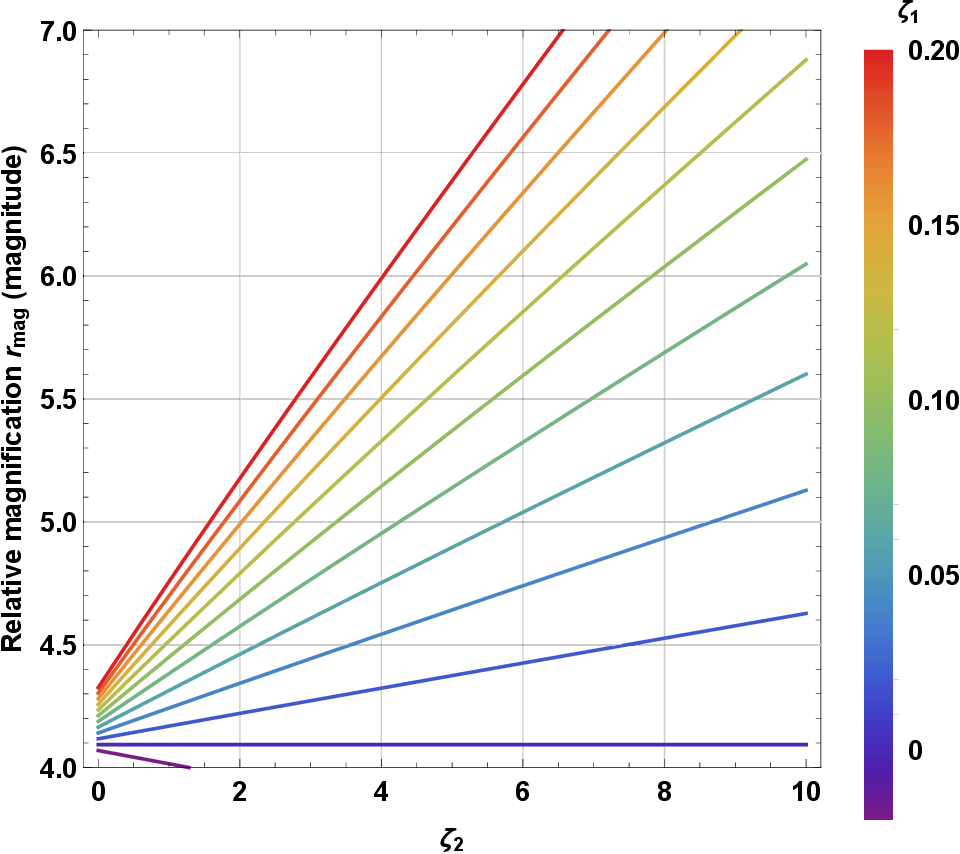}
\caption{The behaviour of the relative magnification $r_{mag}$ with  both the parameters $\zeta_1$ and $\zeta_2$ for the wormhole space-time Model-I.}\label{fig:16}
\end{figure*}

\begin{figure*}
\centering
\includegraphics[width=.45\textwidth]{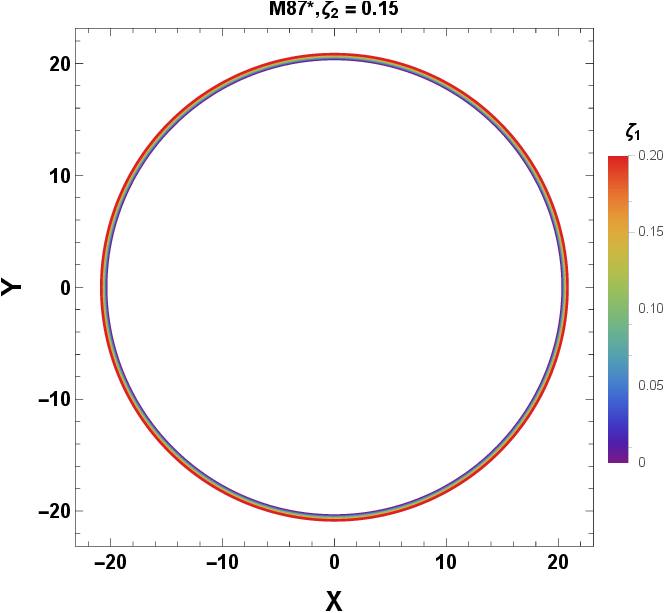}(a)
\qquad
\includegraphics[width=.45\textwidth]{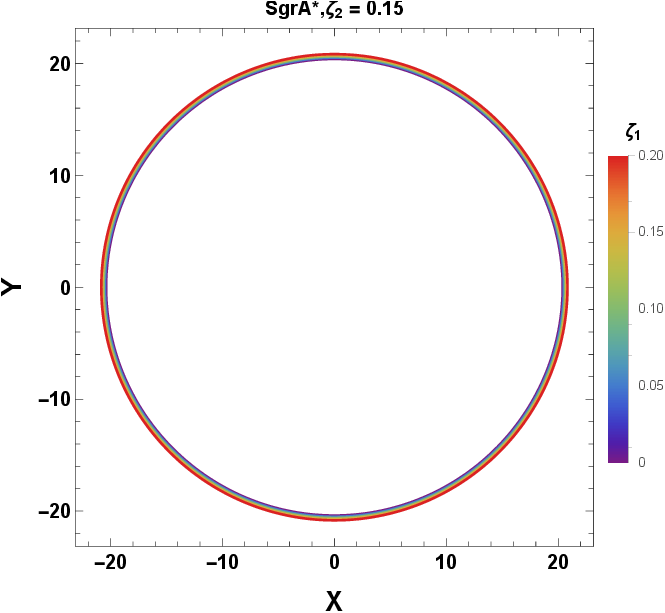}(b)
\caption{The behaviour of the  outermost Einstein's rings $\theta_1^E$ for M87* (left panel) and Sgr A* (right panel)   with  both the parameters $\zeta_1$ and $\zeta_2$ for the wormhole space-time Model-I.}\label{fig:16a}
\end{figure*}
\begin{figure*}
\centering
\includegraphics[width=.45\textwidth]{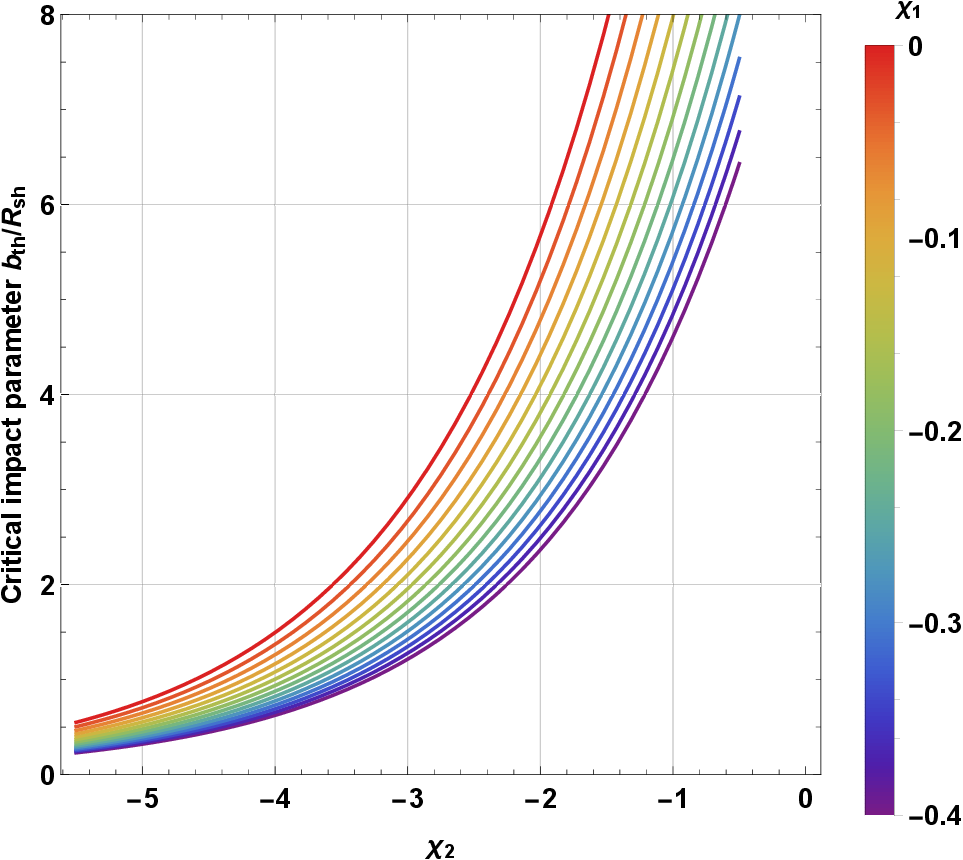}(a)
\qquad
\includegraphics[width=.45\textwidth]{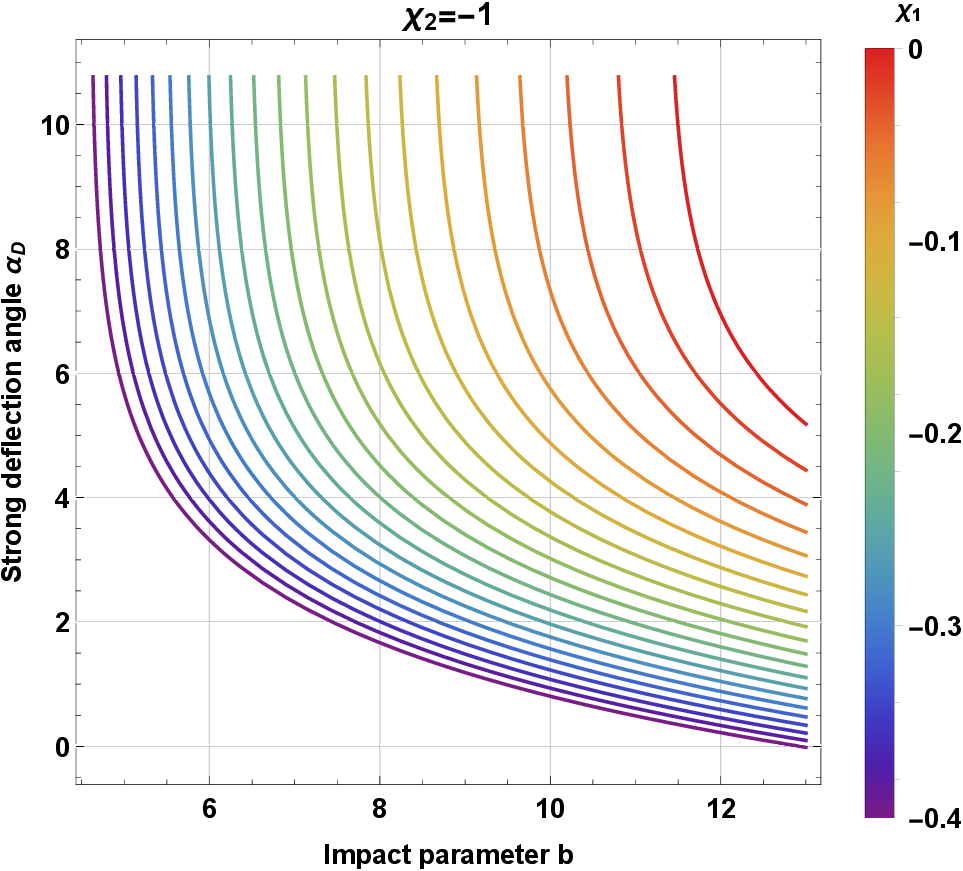}(b)
\caption{The behavior of the critical impact parameter $\mathit{u_{ph}/R_{sh}}$ (left panel) and strong deflection angle $\alpha_D $ (right panel) with  both the parameters $\chi_1$ and $\chi_2$ for the wormhole space-time Model-II.}\label{fig:17}
\end{figure*}
\begin{figure*}
\centering
\includegraphics[width=.45\textwidth]{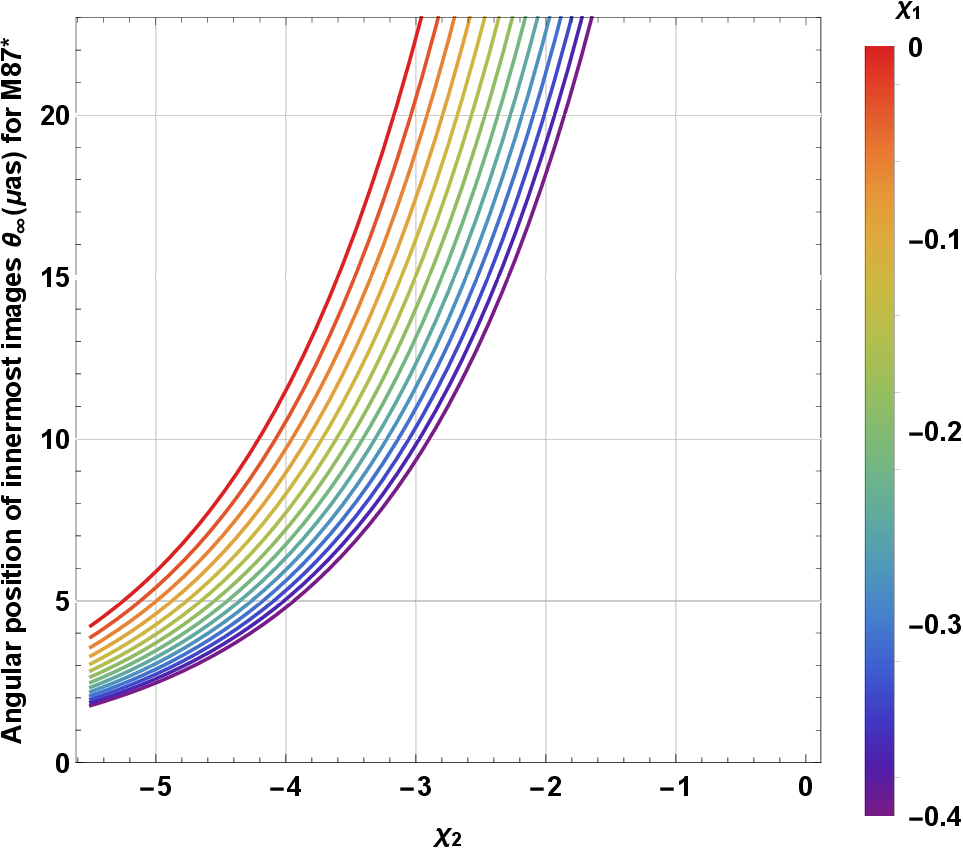}(a)
\qquad
\includegraphics[width=.45\textwidth]{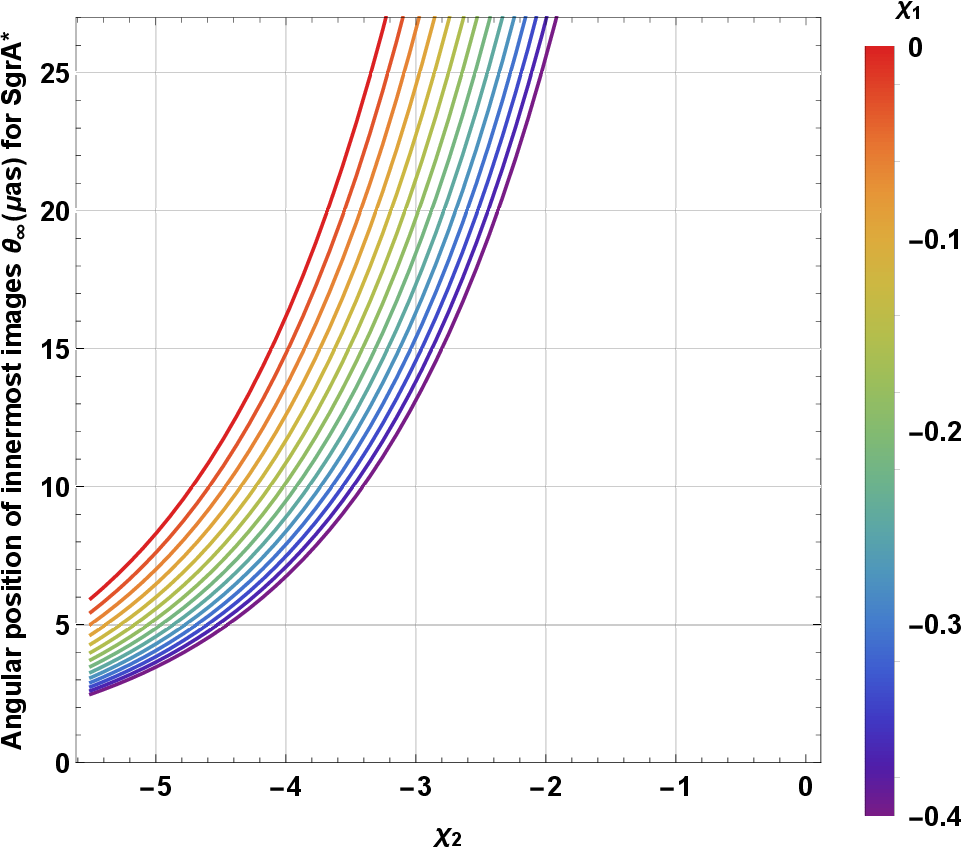}(b)
\caption{The behaviour of the angular position of the innermost image $\theta_{\infty}$ for M87* (left panel) and Sgr A* (right panel)   with  both the parameters $\chi_1$ and $\chi_2$ for the wormhole space-time Model-II.}\label{fig:18}
\end{figure*}

\begin{figure*}
\centering
\includegraphics[width=.45\textwidth]{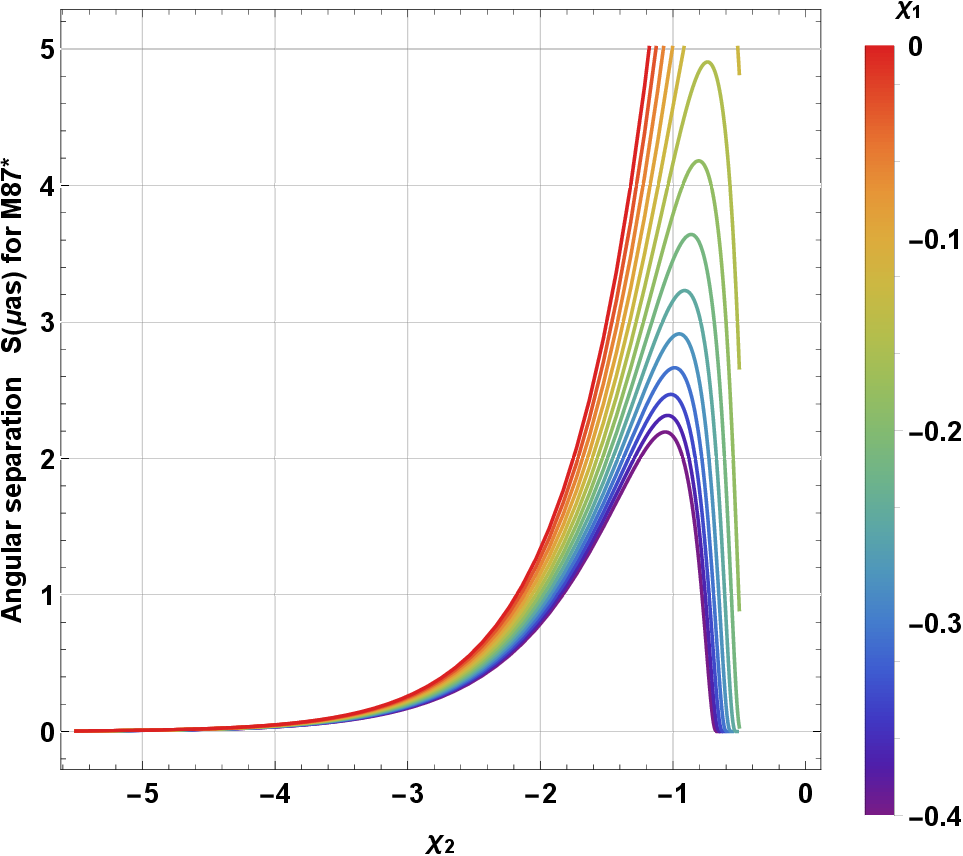}(a)
\qquad
\includegraphics[width=.45\textwidth]{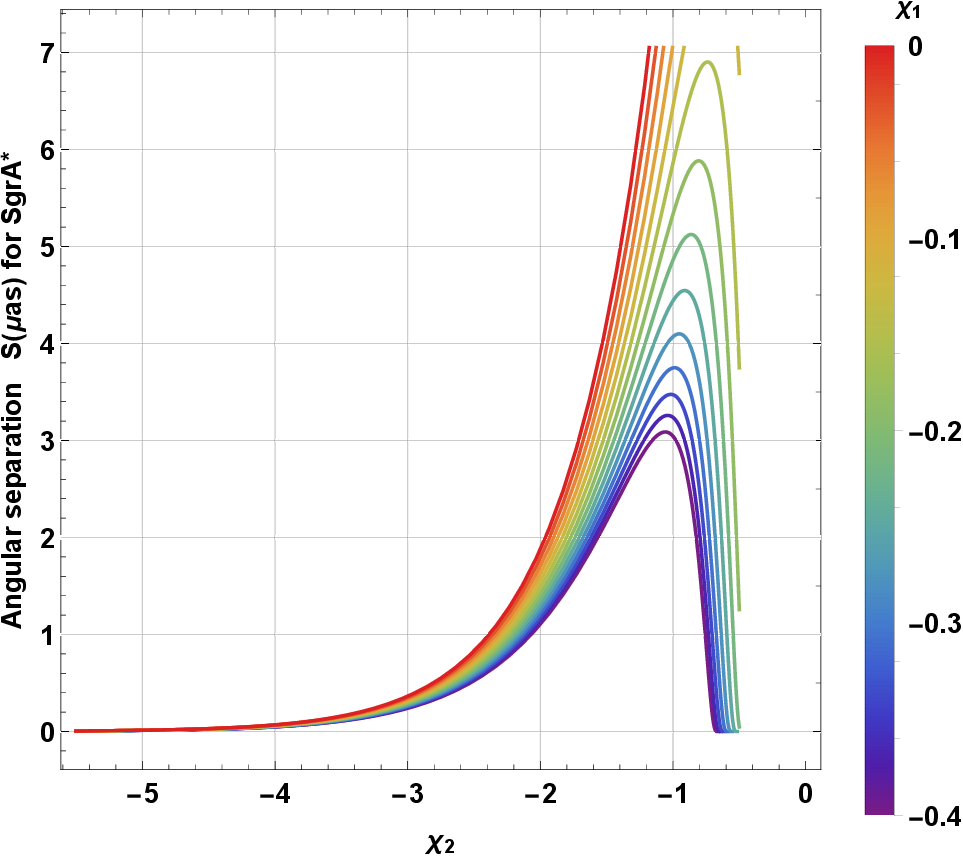}(b)
\caption{The behaviour of the angular separation of the innermost and outermost images $S$ for M87* (left panel) and Sgr A* (right panel)  with  both the parameters $\chi_1$ and $\chi_2$ for the wormhole spacetime Model-II.}\label{fig:19}
\end{figure*}
\begin{figure*}
\centering
\includegraphics[width=.45\textwidth]{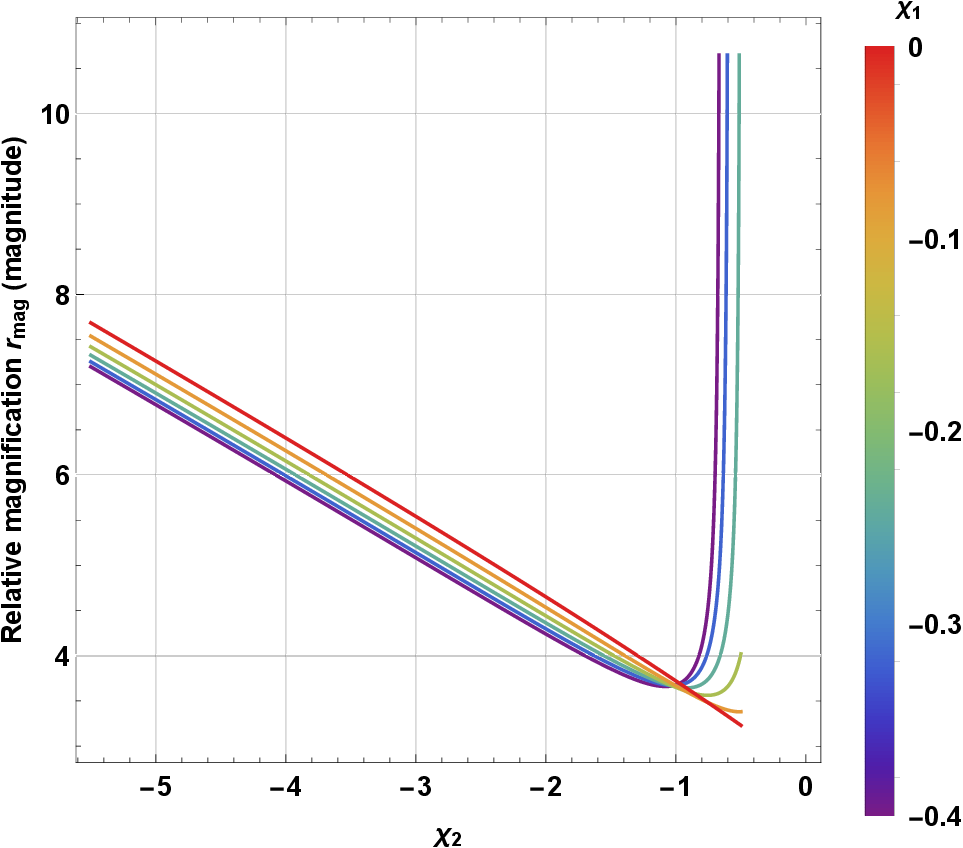}
\caption{The behaviour of the relative magnification $r_{mag}$  with  both the parameters $\chi_1$ and $\chi_2$for the wormhole space-time Model-II.}\label{fig:20}
\end{figure*}

\begin{figure*}
\centering
\includegraphics[width=.45\textwidth]{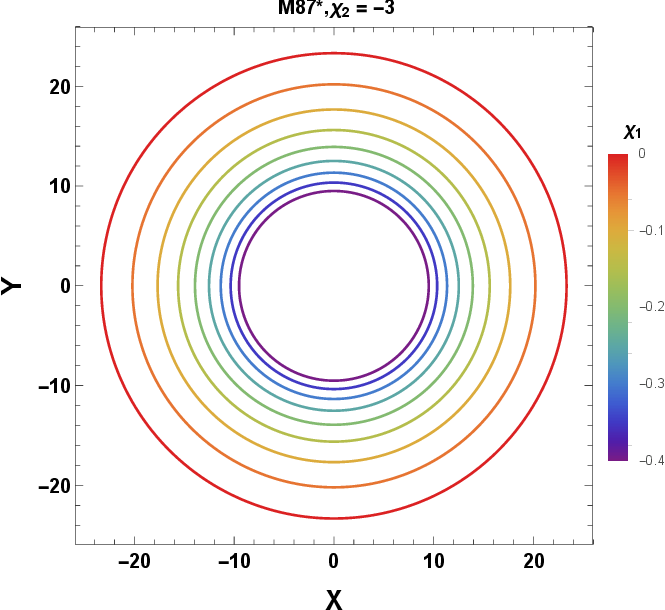}(a)
\qquad
\includegraphics[width=.45\textwidth]{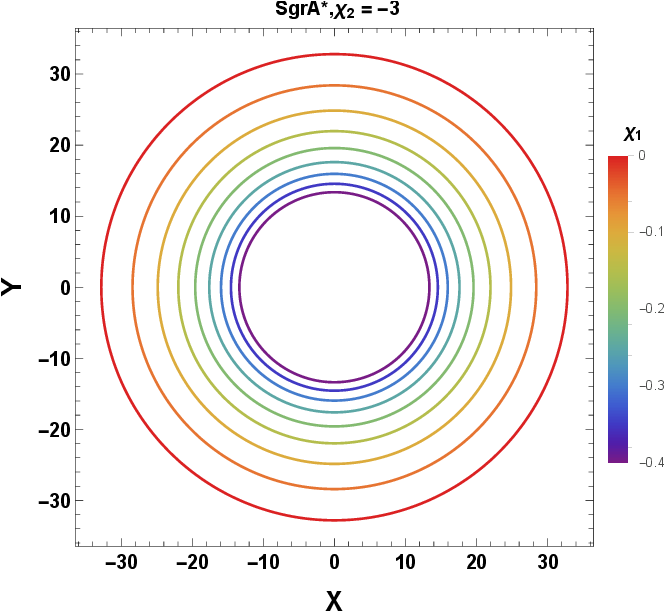}(b)
\caption{The behaviour of the  outermost Einstein's rings $\theta_1^E$ for M87* (left panel) and Sgr A* (right panel)   with  both the parameters $\chi_1$ and $\chi_2$ for the wormhole space-time Model-II.}\label{fig:20a}
\end{figure*}
For the unstable circular photon orbit of radius $r_{ph}$, the conditions for the effective potentials  are$\frac{dV_{eff}}{dr}|_{r_{ph}}=0$ and
$\frac{d^2V_{eff}}{dr^2}|_{r_{ph}}<0$. Thus, the photon sphere radius $r_{ph}$ is the largest real  root of the equation
\begin{equation}\label{37}
    2A(r_{ph})-r_{ph}A^{\prime}(r_{ph})=0
\end{equation}
and the corresponding  critical impact parameter, denoted as $b_{ph}$, is precisely defined by
 \begin{equation}\label{38}
b_{ph} =\frac{r_{ph}}{\sqrt{A(r_{ph})}}
 \end{equation}

When the particle are coming at the  closest distance $r=\tilde{r}_0$ to the wormhole, where $\frac{dr}{d\tau}=0$, one can  define the minimum impact parameter $b_0$ in terms of closest distance $\tilde{r}_0$ \cite{Bozza:2002zj} as
\begin{equation}\label{39}
   b_0=\frac{\tilde{r}_0}{\sqrt{A(\tilde{r}_0)}}
\end{equation}

The  critical impact parameter corresponding to the wormhole throat $b_{th}$ is given by
\begin{equation}\label{40}
b_{th} =\frac{r_{th}}{\sqrt{A(r_{th})}}
\end{equation}

The deflection angle  in the strong field limit for a wormhole spacetime, as a function of  the closest approach distance  $r_0$, can be expressed as: \cite{weinberg1972gravitation,Virbhadra:1998dy,Shaikh:2019jfr}
\begin{equation}\label{41}
\alpha_D(\tilde{r}_0)=I(\tilde{r}_0)-\pi=2\int_{\tilde{r}_0}^\infty \frac{\sqrt{B(r)}dr}{\sqrt{C(r) }\sqrt{ \frac{A(\tilde{r}_0)C(r)}{A(r)C(\tilde{r}_0)}-1} } -\pi.
\end{equation}
 The strong deflection angle $\alpha_D(\tilde{r}_0)$  depends upon the relation between $\tilde{r}_0$ and $r_{th}$ and while $\tilde{r}_0\approx r_{th}$, it is increased.
 So, we consider a new variable z as:
 \begin{equation}\label{42}
 z=1-\frac{\tilde{r}_0}{r}. 
 \end{equation}
For,  $\tilde{r}_0\approx r_{th}$, the strong  deflection angle  can be obtain as
\begin{equation}\label{43}
\alpha_D(b)= -\bar{a}~ log\left(\frac{b}{b_{th}}-1\right) +\bar{b} +\mathcal{O}((b -b_{th})log(b -b_{th})),
\end{equation}
where the  lensing coefficients $\bar{a}$ and  $\bar{b}$ respectively are given by
\begin{equation}\label{44}
    \bar{a}=2\sqrt{\frac{A_{th}}{(C'_{th}A_{th}-A'_{th}C_{th})\bar{B}'_{th}}},
\end{equation}
\begin{equation}\label{45}
    \bar{b}=\bar{a}~log\biggr[2r_{th}\biggr(\frac{C'_{th}}{C_{th}}-\frac{A'_{th}}{A_{th}}\biggr)\biggr]+I_{R}(r_{th})-\pi.
\end{equation}

Here, $``\prime"$ denotes the derivative with respect to $r$, $\bar{B}(r)=1/B(r)$, and $I_{R}(r_{th})$ is formulated in detail in \cite{Bozza:2002zj,Shaikh:2019jfr}.
Keeping all the parameters fixed, the critical impact parameter $b_{th}/R_{sh}$ and the strong deflection angle due to the wormhole throat have been depicted in Fig. (\ref{fig:13}) for Model-I and Fig. (\ref{fig:19}) for Model-II. For Model-I, from Fig. \ref{fig:13}(a), it is seen that the critical impact parameter decreases with $\zeta_2$ for a fixed value of $\zeta_1$ and increases with $\zeta_1$ for a fixed value of $\zeta_2$. From Fig. \ref{fig:13}(b), it is observed that the strong deflection angle decreases with the impact parameter $b$ for a fixed value of both parameter $\zeta_1$  and $\zeta_2$ and it also increases with $\zeta_1$ for a fixed value of $\zeta_2$.For Model-II, for Fig. \ref{fig:19}(a), it is seen that the critical impact parameter increases with $\chi_2$ for a fixed value of $\chi_1$ and also increases with $\chi_1$ for a fixed value of $\chi_2$.From Fig. \ref{fig:19}(b), it is observed that the strong deflection angle decreases with the impact parameter $b$ for a fixed value of both parameter $\chi_1$  and $\chi_2$ and it also increases with $\zeta_1$ for a fixed value of $\chi_2$.

Now, we will discuss some observable quantities of the relativistic images via strong gravitational lensing observations. We consider the case where the observer and the source are far from the wormhole throat. It is also considered that the observer and the source are on the same side of the wormhole throat. The relativistic images due to the wormhole throat by the strong gravitational lensing observation have been discussed in the literature \cite{Shaikh:2019jfr}. It is noted that the wormhole throat acts as a photon sphere. As provided in the literature, the strong deflection angle of photon rays due to the black hole is the same as the strong deflection angle of photon rays due to the wormhole. Therefore, the expression for the angular position of the relativistic images formed for the presented two wormhole cases will be the same as those due to a black hole, and for the $n^{th}$ relativistic images, it is positively expressed by \cite{Shaikh:2019jfr,Bozza:2002zj}.

\begin{equation}\label{46}
\theta _n =  \theta^0 _n - \frac{b_{th} e_n D_{os}(\theta_n^0-\beta)}{\bar{a}D_{ls}D_{ol}},
\end{equation}
where $$ e_n=e^{\frac{\bar{b}-2n\pi}{\bar{a}}},$$
$$\theta^0_n=\frac{ b_{th}(1+e_n)}{D_{ol}}.$$ 
where the angle $\beta$ represents the angular separation between the lens and the source or between the observer and the source. Here, $D_{ol}$, $D_{ls}$, and $D_{os}$ denote the distances from the observer to the lens, from the lens to the source, and from the observer to the source, respectively, such that $D_{os} = D_{ol} + D_{ls}$. Note that the angular image position gradually decreases with $n$, implying that in the image formed outside of the wormhole throat, the first relativistic image is the outermost one, i.e., the image with angular position $\theta_1$ is the outermost one, and the image with angular position $\theta_{\infty}$ is the innermost one.

The magnification  for the  $n$-th relativistic image, is defined as \cite{Bozza:2002zj,Shaikh:2019jfr}
\begin{equation}\label{47}
\mu_n=\frac{e_n b^2_{th}(1+e_n)D_{os}}{\bar{a}\beta D_{ls}d^2_{ol}}.
\end{equation}
Furthermore, we can introduce an additional observable parameter, denoted as the angular separation $S_n$ between the  $n^{th}$ and  $(n+1)^{th}$ images as
\begin{equation}\label{48}
    S_n= \theta_n -\theta_{n+1}.
\end{equation}

It is clear that the first relativistic image is the most luminous, and the magnification diminishes exponentially as we progress to higher image orders, denoted by the variable $n$. Importantly, as $\beta \rightarrow 0$, Eq. \eqref{47} exhibits divergence. Consequently, perfect alignment significantly maximizes the possibility of detecting the images. If we designate $\theta_n$ as the asymptotic position of innermost packed images, then the brightest image, i.e., the outermost image at position $\theta_1$, can be individually distinguished. With the help of the deflection angle  \eqref{43} and the lens equation provided in \cite{Bozza:2001xd}, we have derived three  lensing observable quantities, namely, the angular position of the inner packed images $\theta_{\infty}$, the angular separation $S$ between the outermost image and the remaining inner set of images, and the flux ratio between the outermost  image and the remaining inner packed images can be expressed as
\cite{Shaikh:2019jfr,Kumar:2022fqo,Bozza:2002zj}
\begin{equation}\label{49}
\theta_{\infty}=\frac{b_{th}}{D_{ol}}
\end{equation}
\begin{equation}\label{50}
S= \theta_1-\theta_{\infty}\approx\theta_{\infty}e^\frac{(\bar{b} -2\pi)}{\bar{a}}
\end{equation}
\begin{equation}\label{51}
r_{mag}=\frac{\mu_1}{\Sigma^{\infty}_{n=2}\mu_{n}} \approx \frac{5\pi}{\bar{a}log(10)}
\end{equation}
If the observable quantities $\theta_{\infty}$, $S$, and $r_{mag}$ are obtained from observations, the strong lensing coefficients $\bar{a}$ and $\bar{b}$, and the critical impact parameter $b_{th}$ can be easily obtained using Eqs. \eqref{49}, \eqref{50}, and \eqref{51}. Furthermore, these values can be compared to predictions from theoretical models, allowing for the characterization of a wormhole.
\par 
Another important observable quantity is the Einstein ring, extensively discussed in \cite{Einstein:1936llh}. By simplifying Eq. (\ref{46}) for $\beta=0$, we obtain the angular radius of the $n^{th}$ relativistic image as follows:
\begin{equation}\label{52}
     \theta _n =  \theta^0 _n \biggr(1 - \frac{b_{th} e_n  D_{os}}{\bar{a}D_{ls}D_{ol}}\biggr).
 \end{equation}
In the scenario where the lens is midway between the source
and observer i.e., $D_{os}=2D{ol}$ and taking $D_{ol}>>b_{th}$, the
angular radius of the $n^{th}$ relativistic Einstein ring can be
obtained as
\begin{equation}\label{53}
\theta^E_n=\frac{b_{th}(1+e_n)}{D_{ol}}.
\end{equation}
where $\theta^E_1$ represents the outermost Einstein's ring, which is depicted for Model-I in Fig. (\ref{fig:16a}) and for Model-II in Fig. (\ref{fig:20a}).

\subsection{ Observables findings by the strong  gravitational lensing }
In the present paper, using the lensing coefficients $\bar{a}$,$\bar{b}$ and the critical impact parameter $b_{th}/R_{sh}$($R_{sh}=\frac{2G M}{c^2}$), we have investigated observable strong gravitational lensing due to a wormhole throat in two different wormhole models. Although wormholes are massless, we have considered the wormhole throat size $r_{th}$ to be equal to $3M$ for simplicity. Using the supermassive black holes M87* ( mass  $M= 6.5 \times 10^9 M_{\odot}$ and distance $D_{ol}= 16.8$ Mpc) and  Sgr A* ( mass $M= 4.28 \times 10^6 M_{\odot}$ and distance $D_{ol}=8.32$ kpc), we have obtained strong lensing observables such as the angular position of the innermost images $\theta_{\infty}$, angular separation $S$, and relative magnification $r_{mag}$ of the relativistic images in the context of both two different wormhole models.
\subsubsection{ \bf {Lensing observales for Model-I:}}
Keeping all parameters fixed, the critical impact parameter corresponding to the wormhole throat $b_{th}/R_{sh}$ with the parameters $\zeta_1$ and $\zeta_2$ has been depicted in Fig. \ref{fig:13}(a). It is observed that the critical impact parameter $b_{th}/R_{sh}$ decreases with $\zeta_2$ for a fixed value of $\zeta_1$. Further, it is observed that $b_{th}/R_{sh}$ increases with $\zeta_1$ for a fixed value of $\zeta_2$.The strong deflection angle $\alpha_D(b)$ is plotted with $b=b_{th}+0.00002$, and it monotonically decreases with the impact parameter $b$ for the fixed value of the parameters of $\zeta_1$ and $\zeta_2$. It is also observed that $\alpha_D(b)$ increases with $\zeta_1$ for a fixed value of $\zeta_2$.The behaviour of the observable quantities such as the angular position of the innermost images $\theta_{\infty}$, angular separation between innermost and outermost images $S$, and relative magnification $r_{mag}$ of the relativistic images in the context of M87* and  Sgr A* for wormhole Model-I respectively have been depicted in Figs. (\ref{fig:14}), (\ref{fig:15}), and (\ref{fig:16}). In  Fig. (\ref{fig:14}), it is observed that the angular image position $\theta_{\infty}$ decreases with $\zeta_2$ for a fixed value of $\zeta_1$.In this figure, it is also observed that $\theta_{\infty}$ increases with $\zeta_1$ for a fixed value of $\zeta_2$.
In Fig. (\ref{fig:15}), it is seen that the angular image separation $S$ decreases with $\zeta_2$ for a fixed value of $\zeta_1$. In this figure, it is also seen that $S$ decreases with $\zeta_1$ for a fixed value of $\zeta_2$. From Fig. (\ref{fig:16}), it is observed that the relative magnification $r_{mag}$ increases with a sufficiently larger value of the parameter $\zeta_1$ ($>0$) and the parameter $\zeta_2$, but it decreases with a reasonably small value of the parameter $\zeta_1$ ($<0$) and the parameter $\zeta_2$.
Another important observable quantity, the outermost Einstein ring $\theta^E_1$ for Model-I, in the context of supermassive black holes $M87^{*}$ and $SgrA^{*}$, is depicted in Fig. (\ref{fig:16a}). It is observed that the outermost Einstein ring $\theta^E_1$ increases with $\zeta_1$ for a fixed value of $\zeta_2=0.15$.

\subsubsection{ \bf {Lensing observables for Model-II}}
Here, we briefly review the observable findings due to strong gravitational lensing in the context of wormhole  Model-II.
Keeping all parameters fixed, the critical impact parameter corresponding to the wormhole throat $b_{th}/R_{sh}$ with the parameters $\zeta_1$ and $\zeta_2$ has been depicted in Fig. \ref{fig:17}(a). The critical impact parameter $b_{th}/R_{sh}$ is observed to increase with $\chi_2$ for a fixed value of $\chi_1$. Furthermore, it is observed that $b_{th}/R_{sh}$ increases with $\chi_1$ for a fixed value of $\chi_2$.The strong deflection angle $\alpha_D(b)$ in the context of wormhole  Model-II is plotted with $b=b_{th}+0.00002$ and monotonically decreases with the impact parameter $b$ for the fixed value of the parameters of $\chi_2=-1$ and $\chi_1$. It is also observed that $\alpha_D(b)$ increases with $\chi_1$ for a fixed value of $\chi_2=-1$. Behavior of the observable quantities such as the angular position of the innermost images $\theta_{\infty}$, angular separation between innermost and outermost images $S$, and relative magnification $r_{mag}$ of the relativistic images in the context of M87* and  Sgr A* for wormhole Model-II respectively have been depicted in
Figs. (\ref{fig:18}), (\ref{fig:19}) and (\ref{fig:20}). In Fig. (\ref{fig:18}), It is observed that the angular image position $\theta_{\infty}$ increases with $\chi_2$ for a fixed value of $\chi_1$. In this figure, it is also observed that $\theta_{\infty}$ increases with $\chi_1$ for a fixed value of $\chi_2$.
In Fig. (\ref{fig:19}), it can be observed that the angular image separation $S$ increases with $\chi_2$, reaching its maximum value, and then decreases with $\chi_2$ for a fixed value of $\chi_1$. In this figure, it is also seen that $S$ increases with $\chi_1$ for a fixed value of $\chi_2$. From Fig. (\ref{fig:20}),  It is observed that the relative magnification $r_{mag}$ decreases with $\chi_2$, reaching its minimum value, and then increases with $\chi_2$ for a fixed value of $\chi_1$. In this figure, it is also observed that $r_{mag}$ increases with $\chi_1$ for a fixed value of $\chi_2$.
The observable quantity, the outermost Einstein ring $\theta^E_1$, for Model-II in the context of supermassive black holes $M87^{*}$ and $SgrA^{*}$, is depicted in Fig. (\ref{fig:20a}). It is observed that the outermost Einstein ring $\theta^E_1$ increases with $\chi_1$ for a fixed value of $\chi_2=-3$.

\section{Conclusion} \label{sec7}
\par 
In the current analysis, we have calculated two new embedded wormhole solutions by considering two different red-shift functions. Interestingly, both newly calculated shape functions have satisfied the required conditions for the existence of wormholes. All the essential properties, including Flaring out and flatness properties, have been satisfied for the different choices of involved parameters for both Model-I and Model-II. In order to check the nature of matter, we have checked the regional behavior of all the energy conditions within the scope of different ranges of involved parameters for both newly calculated wormhole solutions. It has been noticed in section IV that, in the maximum region, all the energy conditions have an invalid region or supportive region for the wormhole existence. The violated behavior of all the energy conditions in the maximum region confirms the presence of exotic matter, which is one of the most required components for the existence of these new wormhole solutions.      
We investigate the shadow and strong gravitational lensing due to the wormhole throat for the two new wormhole models, namely Model-I and Model-II. In the present paper, we consider the wormhole throat to act as a photon sphere. Furthermore, to investigate strong gravitational lensing observations, we consider the situation where the observer and light source are on the same side of the wormhole throat. For Model-I, it is found that the radius of the wormhole shadow increases when the parameter $\zeta_1$ increases for the fixed value of the parameter $\zeta_2=1,2$, and it decreases when the parameter $\zeta_2$ increases with the fixed value of the parameter $\zeta_1=0.1$ or $0.2$. Furthermore, for Model II, it is observed that the radius of the wormhole shadow increases with $\chi_1$ for a fixed value of $\chi_2= -1$ or $-2$ and also increases with $\chi_2$ for a fixed value of $\chi_1= -0.1$ or $-0.2$.\\
In order to investigate the effects of the parameters $\zeta_1$ and $\zeta_2$ for Model-I and $\chi_1$ and $\chi_2$ for Model-II on strong gravitational lensing, we have investigated the behavior of critical impact parameter $b_{th}/R_{sh}$ and then using it, we have obtained the strong deflection angle due to the wormhole for Model-I and Model-II. The strong deflection angle $\alpha_D(b)$ with $b=b_{th}+0.00002$  monotonically decreases with the impact parameter $b$ for the fixed value of the parameters of $\zeta_1$ and $\zeta_2$ for Model-I; and for the fixed value of $\chi_1$ and $\chi_2$ for the Model-II. Furthermore, it is observed that strong deflection angle $\alpha_D(b)$  increases with the parameter$\zeta_1$ for the fixed value of the parameter $\zeta_2$ for Model-I and increases with $\chi_1$ for a fixed value of $\chi_2$  for the Model-II. Thus, the wormhole with the presence of parameters $\zeta_1$ and $\zeta_2$ for Model-I and   $\chi_1$ and $\chi_2$ for Model-II intensify the gravitational lensing effects. Therefore, a wormhole with the presence of exotic matter, fluid, etc, around it may be detected more easily through gravitational lensing observation.\\
 We have graphically obtained various strong lensing observables of relativistic images, such as the angular image position of the innermost images $\theta_{\infty}$, the angular separation between outermost and innermost images $S$, the relative magnification $r_{mag}$ and the outermost Einstein ring $\theta^E_1$, in the wormhole spacetime background for both Model-I and Model-II. The behavior of these strong observables is analyzed graphically by considering the example of supermassive black holes, such as $M87^{*}$ and $SgrA^{*}$, in the center of nearby galaxies.
For Model-I, it is observed that the angular image position $\theta_{\infty}$ decreases with $\zeta_2$ for a fixed value of $\zeta_1$. Additionally, it is noted that $\theta_{\infty}$ increases with $\zeta_1$ for a fixed value of $\zeta_2$, while the angular image separation $S$ decreases with $\zeta_2$ for a fixed value of $\zeta_1$. It is also observed that $S$ decreases with $\zeta_1$ for a fixed value of $\zeta_2$. The observable quantity, relative magnification $r_{mag}$, increases with a sufficiently larger value of the parameter $\zeta_1$ ($>0$) and the parameter $\zeta_2$, but it decreases with a sufficiently smaller value of the parameter $\zeta_1$ ($<0$) and the parameter $\zeta_2$.
Another important observable quantity, the outermost Einstein ring $\theta^E_1$ for Model-I, in the context of supermassive black holes $M87^{*}$ and $SgrA^{*}$, has been obtained. It is observed that the outermost Einstein ring $\theta^E_1$ increases with $\zeta_1$ for a fixed value of $\zeta_2$.\\
For Model-II, it is observed that the angular image position $\theta_{\infty}$ increases with $\chi_2$ for a fixed value of $\chi_1$. Additionally, it is noted that $\theta_{\infty}$ increases with $\chi_1$ for a fixed value of $\chi_2$, while the angular image separation $S$ first increases with $\chi_2$, reaching its maximum, then decreases with $\chi_2$ for a fixed value of $\chi_1$. It is also observed that $S$ increases with $\chi_1$ for a fixed value of $\chi_2$. The observable quantity, relative magnification $r_{mag}$, first decreases with the parameter $\chi_2$, reaching its minimum, then decreases with $\chi_2$ for the fixed parameter $\chi_1$, while it increases with $\chi_1$ for the fixed parameter $\chi_2$.
The observable quantity, the outermost Einstein ring $\theta^E_1$, for Model-II in the context of supermassive black holes $M87^{*}$ and $SgrA^{*}$, has been obtained. It is observed that the outermost Einstein ring $\theta^E_1$ increases with $\chi_1$ for a fixed value of $\chi_2$.\\
It is concluded that, keeping all other parameters fixed, it is observed that the parameters $\zeta_1$ and $\zeta_2$ for Model-I and $\chi_1$ and $\chi_2$ for Model-II highly affect the wormhole shadow and gravitational lensing in the strong field limit. Our conclusion is that it may be possible to detect relativistic images, such as Einstein rings, produced by wormholes with throat radii of $r_{th}=3M$. Additionally, current technology enables us to test hypotheses related to astrophysical wormholes.\\

\section*{Data Availability}
No new data were generated in support of this research.

\section*{Conflict of Interest}
The authors declare no conflict of interest.

\section*{Acknowledgement}
N.U.M would like to thank  CSIR, Govt. of
India for providing Senior Research Fellowship (No. 08/003(0141)/2020-EMR-I).
 G. Mustafa is very thankful to Prof. Gao Xianlong from the Department of Physics, Zhejiang Normal University, for his kind support and help during this research. \\

	\end{document}